\documentclass[manuscript]{aastex} % for a referee version
\usepackage{graphicx,epsfig}
\usepackage{natbib}                              
\usepackage{rotating}
\usepackage{txfonts}
\usepackage{longtable,lscape,url}
\usepackage{rotating}

\shorttitle{Long-term Optical Polarization Variability}
\shortauthors{Sorcia et al.}

\begin{document}

\title{LONG-TERM OPTICAL POLARIZATION VARIABILITY OF THE TeV BLAZAR 1ES~1959+650}

\author{Marco Sorcia\altaffilmark{1}, Erika Ben\'itez\altaffilmark{1}, David Hiriart\altaffilmark{2}, Jos\'e M. L\'opez\altaffilmark{2}, Jos\'e I. Cabrera\altaffilmark{1}, Ra\'ul M\'ujica\altaffilmark{3}, Jochen Heidt\altaffilmark{4}, Iv\'an Agudo\altaffilmark{5}, Kari Nilsson\altaffilmark{6} and Michael Mommert\altaffilmark{4,7}} 
\affil{Instituto de Astronom\'ia\\Universidad Nacional Aut\'onoma de M\'exico }

\email{msorcia@astro.unam.mx}

\altaffiltext{1}{Instituto de Astronom\'ia, Universidad Nacional Aut\'onoma de M\'exico, Apdo. 70-264, Mexico D.F., 04510, Mexico}  
\altaffiltext{2}{Instituto de Astronom\'ia, Universidad Nacional Aut\'onoma de M\'exico, Apdo. 810, Ensenada, B.C., 22800, Mexico}
\altaffiltext{3}{Instituto Nacional de Astrof\'isica, \'Optica y Electr\'onica, Apdo. Postal 51 y 216, 72000 Tonantzintla, Puebla, Mexico}
\altaffiltext{4}{ZAH, Landessternwarte Heidelberg, K\"onigstuhl 69117, Heidelberg, Germany}
\altaffiltext{5}{Instituto Nacional de Astrof\'isica de Andaluc\'ia, CSIC, Apartado 3004, 18080, Granada, Spain}
\altaffiltext{6}{Finnish Centre for Astronomy with ESO (FINCA), University of Turku, V\"ais\"al\"antie 20, FI-21500 Piikki\"o, Finland}
\altaffiltext{7}{Deutsches Zentrum f\"ur Luft- und Raumfahrt e.V., Institut f\"ur Planetenforschung, 12489 Berlin-Adlershof, Germany}

\begin{abstract}

A detailed analysis of the optical polarimetric variability of the TeV blazar 1ES 1959+650 from 2007 October 18  to 2011 May 5 is presented.
The source showed a maximum and minimum brightness states in the R-band of 14.08$\pm$0.03 mag and 15.20$\pm$0.03 mag, respectively, with a maximum variation of 1.12 mag, and also a maximum polarization degree of $P=$(12.2$\pm$0.7)\%, with a maximum variation of 10.7\%. From August to November 2009, a correlation between the optical $R$-band flux and the degree of linear polarization was found, with a correlation coefficient $r_{pol}$=0.984$\pm$0.025. The source presented a preferential position angle of optical polarization of $\sim$153$^{\circ}$, with variations of $10\degr$-$50\degr$, that is in agreement with the projected position angle of the parsec scale jet found at 43 GHz. From the Stokes parameters we infer the existence of two optically-thin synchrotron components that contribute to the polarized flux. One of them is stable, with a constant polarization degree of 4\%. Assuming a stationary shock for the variable component, we estimated some parameters associated with the physics of the relativistic jet: the magnetic field, $B\sim$0.06 G, the Doppler factor, $\delta_{0}\sim$23, the viewing angle, $\Phi\sim$2.4$\degr$, and the size of the emission region $r_b\sim$5.6$\times10^{17}$ cm. Our study is consistent with the spine-sheath model to explain the polarimetric variability displayed by this source during our monitoring.

\end{abstract}


\section{Introduction}

Blazars constitute the most extreme subclass of the active galactic nuclei (AGN). They are radio-loud AGN and include the BL Lacertae objects and the Flat Spectrum Radio Quasars \citep{1980ARA&A..18..321A,1990ApJ...354..124I,1995ApJ...443..578M,1997MNRAS.289..136F,1998MNRAS.301..451G}.
BL Lac objects show very weak emission lines with equivalent width of less than 5\,\AA\, or no emission lines at all \citep[e.g.][]{1998MNRAS.301..451G}.
In contrast, flat-spectrum radio quasars (FSRQ) exhibit broad optical emission lines. This suggests that in BL Lac objects accretion disk emission is strongly suppressed \citep{2006IJMPA..21.6015L}. Blazars show strong flux variability, superluminal motion, and a non-thermal continuum extending from radio to TeV $\gamma$-ray regions \citep[e.g][]{1990AJ.....99..769V,2001AIPC..558..275S,2007Ap&SS.307...69B,2011ApJ...726L..13A, 2011ApJ...735L..10A,2010ApJ...710L.126M,2011ApJ...726...43A, 2010Natur.463..919A}.

In the spectral energy distribution (SED) of blazars, the non-thermal continuum emission shows two broad low and
high energy-peaked components. It is widely accepted that the low-energy peak component is produced by synchrotron radiation from a relativistic jet
\citep{1967MNRAS.135..345R,1974ApJ...188..353J,1985ApJ...298..114M}. However, the nature of the high-energy peak component is more controversial.
There are two different approaches to explain its origin. The first one comes from the so-called Leptonic models, which are based on the Inverse Compton (IC)
scattering of soft photons by the same electrons emitting the synchrotron radiation \citep{1981ApJ...243..700K,1985ApJ...298..114M,2009MNRAS.397..985G}. In these models, if the seed photons are provided by the synchrotron radiation emitted at lower energies by the same IC scattering electrons, the models are known as synchrotron self-Compton (SSC,  \citet{1992ApJ...397L...5M,2008MNRAS.385..283C,1985ApJ...298..114M}). If the dominant contribution to the seed photons field for IC comes from regions external to the jet, they are known as External Compton (EC) models. Possible sources of external seed photons include: (i) accretion disk photons entering into the emission region directly \citep{1992A&A...256L..27D,1993ApJ...416..458D}; (ii) photons reprocessed by the clouds within the broad line region (BLR) \citep{1994ApJ...421..153S}; (iii) jet synchrotron emission reprocessed by circumnuclear material located close to the accretion disk and its X-ray corona \citep{1996MNRAS.280...67G}; (iv) infrared emission from a dusty torus surrounding the central engine \citep{2000ApJ...545..107B}.  The second approach used to explain the high-energy peak emission comes from the Hadronic models where it is assumed that the emission is produced by pairs and pion production \citep{1993A&A...269...67M}, as well as by synchrotron radiation from protons, $\pi^{\pm}$ and $\mu^{\pm}$ particles \citep{2000NewA....5..377A,2001APh....15..121M}. In some cases, a hybrid model may be considered when elements of both models might be relevant such as the hadronic synchrotron mirror model  \citep{2007Ap&SS.309...95B}.

A distinctive feature of blazar emission is its high and variable linear polarization in radio and optical bands \citep{1980ARA&A..18..321A,1990ApJ...354..124I}. This property associates the observed emission with beamed synchrotron radiation that is produced by a relativistic jet viewed with a small angle relative to the observer's line of sight \citep{1978bllo.conf..328B,1979ApJ...232...34B,1982MNRAS.199..883B,1995PASP..107..803U,1995ApJ...444..567P,2001AIPC..558..275S,2006IJMPA..21.6015L}. In recent years, there has been an increased interest regarding the optical polarimetric properties of blazars \citep[e.g.][]{2008ApJ...672...40H,2009A&A...508..181D,2009ApJ...697..985D,2010ApJ...710L.126M,2011arXiv1105.0572C,2011A&A...531A..38A,2011PASJ...63..639I}. This is due to the fact that polarization studies can provide useful information on the relative structure and configuration of the magnetic field associated with the relativistic jet. Variations of the position angle of the polarization vector may be associated to variations of the direction of the magnetic field's vector along the line of sight. Also, the degree of optical polarization could be related to the  level of ordering of an initially tangled magnetic field or to the electron energy distribution within the emission region \citep{1980ARA&A..18..321A,2000ApJ...541...66L,2009MNRAS.398.1207D}.

In general, the linear polarization in blazars varies randomly although for some objects a systematic behavior of this quantity is observed during some period of time.
The rotation of the polarization vector might be produced by a magnetic field with a helical structure or by a bending of the jet \citep{2011PASJ...63..489S,2010ApJ...710L.126M}.  
It has been suggested that the polarization random behavior might be produced by several components of the polarized emission \citep{1984MNRAS.211..497H}.  A successful model used to explain the observed polarization in these objects has been proposed by \citet{1996MNRAS.282..788B}, with the inclusion of two components. In such a model, the synchrotron sources of polarized radiation are optically thin and have different polarimetric characteristics. One of the polarized components is stable while the other shows a chaotic behavior. In a study done on the blazar OJ 287 by \citet{1976PASJ...28..117K}, the authors found evidence for two basic components: a steady polarized component, which they suggested is always present, and a randomly variable component, dominating during active phases.

The TeV-blazar 1ES~1959+650 was discovered in the radio band as part of a 4.85 GHz survey performed with the 91~m NRAO Green Bank telescope \citep{1991ApJS...75....1B}.  Systematic observations have been performed with the Very Long Baseline Array \citep[VLBA,][]{2001MNRAS.325.1109B,2003AJ....125.1060R,2004ApJ...600..115P}.  Polarimetric and 43 GHz VLBA observations from 2005-2009 revealed spine-sheath structures in the electric vector position angle (EVPA) and fractional polarization distributions \citep{2010ApJ...723.1150P}. These authors concluded that the blazar 1ES~1959+650 consists of a compact core with a flux density of $\sim$60 mJy and a $\sim$1 milli-arcsecond jet extending to the southeast at a position angle of about $150^{\circ}$. On the other hand, the source was observed also in the optical bands where it displayed large and fast flux variations \citep{1993ApJ...412..541S, 2000A&AS..144..481V, 2004ApJ...601..151K, 2006ApJ...644..742G, 2008ICRC....3.1021H,2008ApJ...679.1029T, 2010ApJ...719L.162B, 2011PASJ...63..639I, 2012Ap&SS.339..339K}. The source is known to be hosted by an elliptical galaxy at $z$ = 0.047 and with $M_{R}=$-23. This host galaxy shows a disk and an absorption dust lane \citep{1999A&A...341..683H}. The SED of 1ES~1959+650 shows its first synchrotron peak at UV-X-ray frequencies, therefore this object is classified as high-peaked BL Lac object (HBL) \citep{2010ApJ...719L.162B}. The mass of the central black hole has been estimated to be $\sim 1.5 \times 10^8 M_{\odot}$ \citep{2002ApJ...569L..35F}. In X-rays, 1ES~1959+650 has been observed with ROSAT and BeppoSAX\citep{2002A&A...383..410B}, with RXTE-ARGOS and XMM-Newton\citep{2002ApJ...571..763G,2004ApJ...601..151K,2006ApJ...644..742G}.  These data showed that the synchrotron peak was in the range of 0.1-0.7 keV, and the overall optical and X-ray spectrum of up to 45 keV is due to synchrotron emission with the peak moving towards high energy frequencies when the flux is high \citep{2003A&A...412..711T}. Also, the first $\gamma$-ray signal at very high energies from 1ES~1959+650 was reported in 1998 by the Seven Telescope Array in Utah \citep{1999ICRC....3..370N}. Subsequently, the source was observed again emitting  high energy $\gamma$-rays \citep[see][]{1999ApJS..123...79H,2003A&A...406L...9A,2003ApJ...583L...9H}. In 2002 this object showed two TeV flares without simultaneous X-ray flares, a behavior sometimes referred to as orphan flares \citep{2004ApJ...601..151K,2005ApJ...621..181D}. Such orphan flares in VHE $\gamma$-rays were not expected within the frame of SSC models \citep{1998MNRAS.301..451G}.  After the detection of TeV emission, the source became a target for different multiwavelength campaigns \citep{2004ApJ...601..151K,2006ApJ...644..742G,2008ApJ...679.1029T,2008ICRC....3.1021H,2010ApJ...719L.162B}. This blazar is listed in The First Catalog of AGN detected with Fermi Large Area Telescope \citep{2010ApJ...715..429A}.

In this paper we report the results obtained from the photopolarimetric monitoring of the TeV-blazar 1ES~1959+650. Our main goal is to establish the long-term optical variability properties of the polarized emission in the R-band.  Variability of the Stoke's parameters obtained from our observations is analyzed in terms of a two-component model. Estimations of some physical parameters that are known to be associated with the kinematics of the relativistic jet were obtained. Also, we present a comparison of the polarization properties found in our study with recent radio maps obtained with the VLBI at 43-GHz on this source by \citet{2010ApJ...723.1150P}. Our work suggests that the observed radiation originates in a region inside the jet where a standing shock is produced. The paper is organized as follows: in section~\ref{Obse} a description of our observations and data reduction process is presented. In section~\ref{Resu} we show our observational results. In section~\ref{Polana} we present the polarimetric data analysis. In section~\ref{Dis} a discussion of our photometric and polarimetric variability results is presented and in section~\ref{Conc} we give a summary of our main results.
Throughout this paper we use a standard cosmology with $H_{0}$\,=\,71km s$^{-1}$ Mpc$^{-1}$ , $\Omega_m$ = 0.27, and  $\Omega_{\Lambda}$=0.73.

\section{Observations and Data Reduction}
\label{Obse}

The observations were carried out with the 0.84~m- f/15 Ritchey-Chr\'etien telescope at the Observatorio Astron\'omico Nacional of San Pedro M\'artir (OAN-SPM) in Baja California, 
M\'exico. We used the instrument Polima which is a direct image polarimeter \citep[see][and references therein]{2011RMxAC..40..131S}.  All observations were taken as a part of a support program dedicated to the optical R-band monitoring of 37 blazars in the framework of the GLAST-AGILE Support Program (GASP) of the Whole Earth Blazar Telescope (WEBT) \citep[e.g.][]{2010ASPC..427..308V}.\footnote{A detailed description of our photopolarimetric monitoring program on TeV Blazars can be found in http://www.astrossp.unam.mx/blazars} Polima consists of a rotating Glan-Taylor prism driven by a stepper motor with an accuracy of 0.1$^\circ$. The Polima's prism specifications show a transmittance of $90\%$, extinction ratio between the two states of polarization of 5$\times 10^{-5}$,  and a wavelength range of 215-2300~nm. Polima has a clean (unvignetted) field of view of (3$\times$3)\arcmin in the sky plane. The telescope has an equatorial mount and its pointing angle spans from declinations between +70 and -45 degrees, and hour angles between +5 and -5 hours. We used three different CCD cameras during the monitored period with formats of (1$\times$1)k in two of them, and one with (2$\times$2)k. Therefore, pixel sizes go from 13.5 to 24\,$\mu$m and plate scales from 0.22 to 0.39~arc-sec pixel$^{-1}$. Due to the small pixel size of the CCDs, and to decrease the integration and readout times, a binning mode of $2\times 2$ was used in all the observing runs. The exposure time was $240$~s per image for 1ES~1959+650. Polima is a single-beam device with a very slow modulation and quite sensitive to the sky noise level. The errors obtained with this instrument could be larger when the sky noise is high since it will dominate over the instrumental errors, so photometric conditions are required for an accurate polarimetry. The optical polarimetric monitoring of the blazar 1ES~1959+650 was carried out from 2007 October 18 JD(2454392) to 2011 May 5 JD(2455687). During this period of time, we carried out 25 observing runs of seven nights per run, centered on the new moon phase when the object was visible. In total we collected 106 data points.

Four images with the $R$-band  filter and relative position angles were taken.  The sequence of 0$^\circ$, 90$^\circ$, 45$^\circ$, and 135$^\circ$ for the relative prism position angle was used to reduce the sky variations influence to calculate both linear polarization and flux from the object. Observations of nearby blank sky regions were obtained at different prism orientations using the same on-chip exposure times and on-source observations. Dark-current frames were not taken since the detectors were always operating at cryogenic temperatures. Flat fields were taken at the four polarizer positions in the $R$ filter at dusk and dawn.  Bias frames were also taken at the middle of the night. For each prism position, bias frames were subtracted from the flat field images.  Then, flat field images were combined to obtain an average flat field image for each prism position. The bias frame was also subtracted from all object images and the resulting frames were multiplied by the mean value of the combined bias-corrected flats. Finally, the object image was divided by the combined flat field. Data were reduced with a pipeline especially written for our monitoring program (Hiriart 2013, in prep). Polarimetric calibrations were done using the polarized standard stars ViCyg12 and HD155197 and the unpolarized standard stars GD319 and BD+332642 \citep{1992AJ....104.1563S}. The instrumental polarization found is $0.6\pm0.5\%$ and the zero position angle at $(-90\pm2)^{\circ}$.
Photometric $R$-band magnitudes were determined from two orthogonal measurements: $f_1=f(0^\circ)+f(90^\circ)$ and $f_2=f(45^\circ)+f(135^\circ)$ where $f(x)$ is the flux of the object (or standard star) obtained at polarizer position $x$. The instrumental flux is the average of these two fluxes, from which instrumental magnitudes for each object were obtained. The magnitudes were measured using the aperture photometry technique. Then, we calculated the object's $R$-band magnitudes using comparison star~\#2, located $\sim$2.1~\arcmin away from the target. The calibrated magnitude of the comparison star~\#2 in the $R$-band used is 12.53$\pm$0.02 mag, from \citet{1998A&AS..130..305V}.

We calculated the normalized Stokes parameters $q$ and $u$ for each object as:

\begin{equation}
q = \frac{f(0^\circ) - f(90^\circ)}{f(0^\circ) + f(90^\circ)} \; ,
\label{eq_q}
\end{equation}
and
\begin{equation}
u = \frac{f(45^\circ) - f(135^\circ)}{f(45^\circ) + f(135^\circ)} \; .
\label{eq_u}
\end{equation}

The $q$ and $u$ polarization parameters are related to the fractional polarization $p$ and the position angle of polarization $\theta$ by:
\begin{equation}
p=\sqrt{q^2 + u^2}  \; ,
\label{polarization}
\end{equation}
and
\begin{equation}
\theta = \frac{1}{2} \arctan   ( \frac{u}{q})  \; .
\label{angle}
\end{equation}

Defining the fractional polarization in this way, we are assuming that the circular polarization is negligible (see ~\citet{1988ApJ...332..678J}).

\subsection{Error calculations}

From equations (\ref{eq_q}) and (\ref{eq_u}), we propagate the errors for $u$ and $q$ to obtain:

\begin{equation}
\sigma_u = \sqrt{[\frac{2f_{135}}{(f_{45}+f_{135})^2}\sigma_{f_{45}} ]^2+
  [\frac{2f_{45}}{(f_{45}+f_{135})^2} \sigma_{f_{135}}]^2} \; ,
\label{sigmau}
\end{equation}
and
\begin{equation}
\sigma_q = \sqrt{[\frac{2f_{90}}{(f_{0}+f_{90})^2}\sigma_{f_{0}} ]^2+
  [\frac{2f_{0}}{(f_{0}+f_{90})^2} \sigma_{f_{90}}]^2} \; ,
\label{sigmaq}
\end{equation}
where $\sigma_f$ is the error in the instrumental flux \citep[see][]{1991PASP..103..122N} and $f_x$ is the
instrumental flux at the orientation $x$ of the polarization analyzer.

Similarly, the errors in the polarization degree, $\epsilon_P$, and
polarization position angle, $\epsilon_\theta$, follow from
equations~(\ref{polarization}) and (\ref{angle}), we obtain:

\begin{equation}
\epsilon_P =\sqrt{\sigma_u^2+\sigma_q^2} \;,
\label{pol}
\end{equation}
and
\begin{equation}
\epsilon_\theta = \frac{1}{2p}\sqrt{(q \sigma_u )^2+ (u \sigma_q)^2} \; ,
\label{sigmatheta}
\end{equation}
where $\sigma_u$ and $\sigma_q$ are given by eqs.~(\ref{sigmau})~ and
(\ref{sigmaq}), respectively, and $p$ is the measured polarization.

\section{Observational Results}
\label{Resu}
\subsection{Global variability properties}
\label{Resu1}
 
The $R$-band magnitudes were converted into apparent fluxes using the expression: $F_{obs}=K_0 \times10^{-0.4m_R}$, with $K_0= 3.08\times 10^6$ mJy, for an effective wavelength of  $\lambda = 640 \,$nm. The photometry was done with an aperture of 3~\arcsec \ in all runs, and the obtained images have an average FWHM (Full Width and Half Maximum)  $\sim$3 \arcsec. The fluxes were corrected for the contribution of the host galaxy, subtracting 0.84$\pm$0.02 mJy according to \citet{2007A&A...475..199N}.

We have to consider the ambiguity of $180^{\circ}$ in the polarization angle.  For this purpose we corrected the polarization angle assuming that the differences between the polarization angle of temporal adjacent data should be less than $90^{\circ}$. We defined this difference as:
\begin{equation}
 |\Delta\theta_{n}|=|\theta_{n+1}-\theta_{n}| - \sqrt{\sigma(\theta_{n+1})^2+\sigma(\theta_{n})^2}, 
\end{equation}
 where $\theta_{n+1}$ and $\theta_{n}$ are the $n+1$ and n-th polarization angles and $\sigma(\theta_{n+1})$ and $\sigma(\theta_{n})$ their errors. If $|\Delta\theta_{n}|\leq90^{\circ}$, no correction is needed. If $\Delta\theta_{n}\,<\,-90^{\circ}$, we add $180^{\circ}$ to $\theta_{n+1}$. If $\Delta\theta_{n}\,>\,90^{\circ}$, we add $-180^{\circ}$ to $\theta_{n+1}$ \citep{2011PASJ...63..489S}. 

Therefore, all data corrected by the contribution of the host galaxy and the 180$\degr$ ambiguity are presented in  Table~\ref{tbl-1}. There, we show in column (1) the Julian Date, in columns (2) and (3) the polarization degree and their errors, in columns (4) and (5) the polarization position angle and their errors, in columns (6) and (7) the $R$-band magnitudes and their errors. The $R$-band fluxes and their errors are given in columns (8) and (9). Figure~\ref{fig1} shows the $R$-band light curve, the polarization degree $p$ and the position angle $\theta$ obtained for all 25 observing runs in $\sim$3.6yr. For clarity, the entire period of observations has been divided into three main cycles: Cycle I:  from 2008 May 04 to 2008 Dec 03 (JD 2454591-2454804). Cycle II: from 2009 Apr 22 to 2009 Nov 18 (JD 2454944-2455154). Cycle III: from 2010 May 08 to 2011 May 05 (JD 2455325-2455687). These cycles are marked with dashed vertical lines in Figure~\ref{fig1} and will be discussed in more detail in the next paragraphs. 

 To perform the statistical analysis on our data, nightly average values and the corresponding standard deviations were estimated. In order to estimate the variability in flux, polarization  degree and polarization position angle, a $\chi^2$-test was carried out. 
The amplitude of the variations $Y\%$ was obtained following \citet{1996A&A...305...42H}, but their relations were applied to flux densities instead of magnitude differences.
Therefore,
\begin{equation}
  Y(\%) = \frac{100}{\cal h S i}\sqrt{(S_{max}-S_{min})^2-2\sigma^2_c} \;\; ,
\end{equation}
where $S_{max}$ and $S_{min}$ are the maximum and minimum values of the flux density, respectively. ${\cal hSi}$ is the mean value, and $\sigma^2_c = \sigma^2_{max}+\sigma^2_{min}$. The variability is described by the fluctuation index $\mu$ defined by 
\begin{equation}
    \mu = 100\frac{\sigma_S}{\cal hSi}\% \; ,
\end{equation}
and the fractional variability index of the source $\cal F$ obtained from the individual nights:
\begin{equation}
    {\cal F} = \frac{S_{max}-S_{min}}{S_{max}+S_{min}} \; .
\end{equation}

Table~\ref{tbl-2} shows the results obtained from the statistical analysis. In column (1) the corresponding cycle is shown, in column (2) the variable parameters, from columns (3) to (10) we present the average value for each variable parameter, the maximum and minimum value observed,  the maximum variation obtained $\Delta_{max}$, the variability amplitude Y(\%), the variability index $\mu$(\%), the variability fraction $\cal F$ and the statistic $\chi^{2}$, respectively. 

We found for 1ES~1959+650 an average flux of $\cal h$$F$$\cal i$\,=\,4.38$\pm$1.02 mJy, with a variation of $Y_{F}=105.25\%$, corresponding to $\Delta$F$\,=\,$4.62\, mJy; the fluctuation index is $\mu_{F}\,=\,23.28\%$ and a fractional variability index is 
${\cal F}_{F}$\,=\,0.47. The maximum and minimum brightness are 14.08$\pm$0.03 mag (7.19$\pm$0.17 mJy) and 15.20$\pm$0.03 mag (2.57$\pm$0.06 mJy), respectively (considering all cycles). 

We have estimated the minimum flux variability timescale using the definition proposed by \citet{1974ApJ...193...43B}:
\begin{equation}
 \tau=dt/\ln(F_1/F_2) \;\; ,
\end{equation}
where $dt$ is the time interval between flux measurements $F_1$ and $F_2$, with $F_1>F_2$. We have calculated all possible timescales 
$\tau_{ij}$ for any pair of observations for which $\mid F_i-F_j\mid >\sigma_{F_i}+\sigma_{F_j}$ at frequency $\nu$. The minimum timescale
is obtained when:
\begin{equation}
  \tau_{\nu}=\mbox{min}\{\tau_{ij,\nu}\} \; , 
\end{equation}
where $i=1,...,N-1; j=i+1,...,N,$ and $N$ is the number of observations. The uncertainties associated to $\tau_{\nu}$ were obtained through the errors in the flux measurements. The minimum flux variability timescale obtained from our data in the $R$-band is $\tau_R$=$t_{min}$=9.74$\pm$1.17 d. This result is in agreement with ~\citet{2004ApJ...601..151K} and \citet{2008ApJ...679.1029T}. 

Using the results of the observational runs presented here, we can quantify the duty cycle (DC) following \citet{1999A&AS..135..477R}:

\begin{equation}
DC=100\frac{\sum_{i=1}^{n}\,N_i(1/\Delta t_i)}{\sum_{i=1}^{n}\,(1/\Delta t_i)}\%,
\end{equation}
where $n$ is the number of nights, $N_i$ is equal to 0 or 1, depending on whether or not variations were detected in  $\Delta t_i = \Delta t_{i,obs}(1+z)^{-1}$,  where $\Delta t_i$ is the time interval at the source rest frame, and $\Delta t_{i,obs}$ is the time interval measured between each run (at the observer's rest frame). We obtained a duty-cycle $DC=38.7\%$ in the $R$-band flux for variations that are above $2\sigma$, with $\sigma =0.27$ mJy. 

\subsection{Photometric variability}

In addition to our main three Cycles, we include in Table~\ref{tbl-2} the results obtained considering the entire data set. From these data, a maximum of $R=$14.08 mag is observed, and a variation of $\Delta\,m_{R}=$1.12 mag in $\Delta t=$329 d (or 0.9 yr) is found. Therefore, during our monitoring period the source shows a maximum scale brightness variation in time scales of $\Delta t\sim$ months.

In Cycle I, the object showed a discrete variability with a relative flux maximum of 4.46$\pm$0.07 mJy (14.60$\pm$0.02 mag) in 2008 May 04 (JD 2454591), and a relative minimum of 3.33$\pm$0.06 mJy (14.91$\pm$0.02 mag) in 2008 Dec 03 (JD 2454804). There is a moderate rise in flux between 2008 June 06 (JD 2454624) and 2008 October 26 (JD 2454766) from 3.74$\pm$0.06 mJy to 4.32$\pm$0.06 mJy in 142 d, at a rate of 0.004 mJy/d. 

In Cycle II the object has been active for 148~d from 2009 April to 2009 September (JD 2454944-2455092). The peak flux of the object was 7.19$\pm$0.17 mJy (14.08$\pm$0.02 mag) at 2009 June 19 (JD 2455002). In this cycle we found a maximum brightness which is also a maximum value observed for the entire period of observations. At the onset of the activity, 2009 April 22 (JD 2454944), the flux was 5.66$\pm$0.08 mJy (14.34$\pm$0.01 mag), then increased to 7.19$\pm$0.17 mJy in 58 d (with a rising rate of 0.026 mJy/d). In 2009 September 17 (JD 2455092) it fell to 4.92$\pm$0.07 mJy (14.49$\pm$0.01 mag) in about 90 d (with a decay rate of  $\sim$ 0.025 mJy/d).

Finally, in Cycle III the source showed a constant increasing flux from 2.57$\pm$0.06 mJy (15.20$\pm$0.02 mag) in 2010 May 14 (JD 2455331) to 5.96$\pm$0.08 mJy (14.28$\pm$0.01 mag) in 2010 November 04 (JD 2455505) in 174 d (rising rate $\sim$0.020 mJy/d). 

\subsection{Polarimetric variability}
\label{Polvar}

\subsubsection{Polarization degree variability}
To establish a possible correlation between the polarization degree and the $R$-band fluxes, a Pearson's correlation coefficient was calculated 
($r_{pol}$). This coefficient was tested through the Student's $t$-test. Using all data, we found that there is no correlation between the $R$-band flux and the polarization degree. The polarization  degree showed a random variability behavior, and a maximum and a minimum of 12.2\% (JD 2454656) and 1.5\% (JD 2455685), respectively. So, the maximum variability observed was $\Delta P$=10.7 \%, in $\Delta t = $1029d (or 2.8 yr).

From the light curve obtained for each cycle (Figure~\ref{fig1}), the Cycle II was split into two periods that we called Cycle IIa from 2009 Apr 22 to 2009 Jun 25, and Cycle IIb from 2009 Aug 14 to 2009 Nov 18 (see Figure~\ref{fig2}). We cannot see any correlation between the $R$-band brightness and polarization degree in Cycle IIa. However, for Cycle IIb we found a strong correlation between these two parameters, with a Pearson's correlation coefficient of $r_{pol}=0.984 \pm 0.025$, see Figure~\ref{fig3}. 

The maximum and minimum polarization degrees for each cycle are shown in the Table~\ref{tbl-2}. In Cycle I, the maximum variability observed of the polarization degree is $\Delta P=8.9\%$ in $\Delta\,t=34$d;  in Cycle IIa, $\Delta P= 3.5\%$ in $\Delta t$ = 30d; in Cycle IIb, $\Delta P= 5.0\%$ in $\Delta t$ = 90d; and in Cycle~III, $\Delta P= 6.9\%$ in $\Delta t$=29d.

\subsubsection{Position angle variability}

In general our data show that there is not a clear correlation between the polarization angle and the $R$-band flux. However, the polarization angle 
shows a general trend to remain around an average value. 
We have looked for a possible correlation between the polarization degree and the position angle using the entire data set, see Figure~\ref{fig4}. This plot shows that $\theta$ has an average value of 153 $\pm$ 16$^\circ$, with maximum variations of $\sim$ $50^{\circ}$ around the average value. This suggests, that the position angle has a preferential position of $\sim150^{\circ}$, independently of the polarization  degree. The preferred position angle of optical polarization is in good agreement with the projected position angle of the parsec scale jet found by  \citet{2010ApJ...723.1150P} from 43 GHz images. We did not observe any significant rotation of the position angle during the entire period of observations. 

In Cycle~I, the polarization angle rises from $135^{\circ}$ (JD 2454594) to $181^{\circ}$ (JD 2454681) at a rate of 0.53$^\circ$ per day, and falls to 132$^\circ$ (JD 2454707) with a rate of 1.88$^\circ$ per day; i.e. variations of 10$\degr$- 30$\degr$ per run.  When the magnitude falls to 14.75 mag (JD 2454773) the polarization angle rises to $186^{\circ}$. So, the polarization angle shows two oscillations when the magnitude increases.  In Cycle~II, the polarization angle is close to a constant value, in the period of maximum activity of the source.  However, there is a slight change in the polarization angle from 146$^{\circ}$ to 164$^{\circ}$, i.e. $\Delta \theta\sim$18$^{\circ}$, when the magnitude falls from 14.44 mag to 14.72 mag (JD 2455059-2455154).  In Cycle~III, we see a steady increase in brightness while the polarization angle $\theta$ changes from 157$^{\circ}$ to 203$^{\circ}$. It is worth to note that in this cycle, the amplitude variations of the polarization angle increase from 10$^{\circ}$ to 50$^{\circ}$ over the average value, as it does the brightness of the source.

\section{Polarimetric Analysis}
\label{Polana}

The study of the variability properties of the polarized emission in blazars allowed us to derive some properties associated with the variable source. In particular, for 1ES1959+650 we found from our observations that the polarization degree shows in general a random behavior, except in Cycle IIb where it follows the $R$-band flux. Also, we found a lack of correlation between the polarization angle and the $R$-band flux. These results could be explained in terms of the presence of one or more variable polarization components. 

A possible scenario that could explain our observational results is the ``spine-sheath'' model \citep{2005A&A...432..401G}. This model was proposed for TeV BL Lacs and invokes  a slower external flow surrounding a fast spine. The slow layer (sheath) could be the result of the interaction of the jet walls with the ambient medium. Another model that can be considered for our study is the two component model \citep[see][]{2008ICRC....3.1021H,2002MNRAS.336..721K}. This model proposes the presence of two emission regions: a steady, quasi-stationary component, and another rapidly variable SSC component. Also, we want to mention that a multi-component scenario for 1ES~1959+650 has been proposed to explain the observational differences found in the radio, optical, X-ray, and $\gamma$-ray bands \citep{2004ApJ...601..151K, 2008ApJ...678...64P, 2008ApJ...679.1029T}. 

\subsection{Parameters of the stable component}

In terms of the normalized Stokes parameters, the absolute Stokes vector is defined as ($Q,U$)=($q,u$)$I$. To identify the presence of a stable polarized component, we have used the method suggested by \citet{1985ApJ...290..627J}. In this work, the authors proposed that if the observed average values ($\cal h$$Q$$\cal i$, $\cal h$$U$$\cal i$) in the absolute Stokes parameters plane $Q$-$U$ deviate significantly from the origin, then a stable or a long-term  $(\sim$\,months-years) variable polarization component is present. 

For the case of a two-component model, we define the average values of $Q$ and $U$ as the stationary polarization component. These averages were calculated iteratively. First, we calculated the average of $Q$ (or $U$) using all data, then the obtained value was recalculated again after discarding the outliers ($>3\sigma$). The iteration continued until no outliers remained \citep{2011PASJ...63..639I}. Figure~\ref{fig5} shows the $Q$-$U$ plane for the stable component.The obtained average values of the absolute Stokes parameters $\langle Q \rangle$ and $\langle U \rangle$ are -0.10$\pm$0.01 mJy and 0.16$\pm$0.02 mJy, respectively. These average values appear as a black-filled star in the $Q$-$U$ plane plot, and correspond to a stable component with constant polarization degree $P_c= 4.1\pm0.5\%$ and polarization angle $\Theta_c=$151\degr$\pm$13$\degr$. The constant polarization degree has a dispersion $\sigma_{P_{c}}=$2.1\%.

\subsection{Parameters of the variable component}

To find the origin of the variability behavior in Cycle II, we used the method proposed by \citet{2008ApJ...672...40H}. Therefore, it is assumed that the variability within certain time interval is due to a single variable component. If the variability is caused only by its flux variations, while the relative Stokes parameter $q$ and $u$ remain unchanged, then in the space of the absolute Stokes parameters $\{I,Q,U\}$ the observational points must lie on straight lines. The slopes of these lines are the relative Stokes parameters of the variable component.

To estimate the variable component parameters, we looked for a possible linear relation between $Q$ vs $I$ and $U$ vs $I$ for the three relevant Cycles. For Cycles I and III no linear correlation between these parameters was found; rather, they appear to be randomly related. 
In contrast, for Cycles IIa and IIb (see Figure~\ref{fig6}), our data show a linear tendency between these parameters. We made a least square fit to the data in order to find the slopes and the linear correlation coefficients $r_q$ and $r_u$.  The correlation coefficients obtained for these parameters in Cycle IIa are $r_q$=0.913 and $r_u$= -0.902, and the slopes $m_q$=0.228$\pm$0.028, $m_u$=-0.408$\pm$0.054, respectively.  The correlation coefficients for Cycle IIb are $r_q$=0.872 and $r_u$=0.980, and slopes  $m_q$=0.304$\pm$0.044, $m_u$=0.275$\pm$0.015, respectively. 
The Stokes parameters for the variable components $q_{var}$ and $u_{var}$ found are given in Table~\ref {tbl-3}. In this table, columns (2) to (7) show the parameters $q_{var}$, $r_{q}$, $u_{var}$, $r_{u}$, $p_{var}$ and $\theta_{var}$, respectively. It is important to note, that $Q$ vs $I$ appear correlated during Cycle IIa and Cycle IIb. On the other hand, $U$ vs $I$ are anti-correlated during Cycle IIa and correlated in Cycle IIb. The polarization degree found for the variable component in Cycle IIa is $p_{var}\,=\,(46.8\pm4.9$)\%, with a polarization angle $\theta_{var}\,=\,150\degr\pm4\degr$. For Cycle IIb, the polarization  degree is $p_{var}\,=\,(41.0\pm3.4$)\% and the polarization angle is $\theta_{var}=21\degr\pm4\degr$.

\section{DISCUSSION}
\label{Dis}
\subsection{Two-component Model}

The $Q$-$U$ plane built from our data shows that the average values are shifted from the origin (Figure~\ref{fig5}). Therefore, we can infer the presence of a constant or stable component that we assume associated with the relativistic jet, and also a variable component that can be related to the propagation of the shock. The observed polarization would be the result of the overlap of these two optically thin synchrotron components, according to the following expressions \citep[see][]{1984MNRAS.211..497H}:

\begin{equation}
  p^2=\frac{p_{cons}^2+p_{var}^2I^2_{v/c}+2p_{var}p_{cons}I_{v/c}\cos2\xi}{(1+I_{v/c})^2}
  \; \; ,
\end{equation}
and
\begin{equation}
\tan2\theta=\frac{p_{cons}\sin2\theta_{cons}+p_{var}I_{v/c}\sin2\theta_{var}}{p_{cons}\cos2\theta_{cons}+p_{var}I_{v/c}\cos2\theta_{var}}
\;\; ,
\end{equation}
where $p$ is the observed polarization degree and $\theta$ the observed position angle. Here, $\xi=\theta_{cons}-\theta_{var}$ and $I_{v/c}$ is the flux ratio between the variable to the constant component.

To solve these equations, we used the previously found parameters $P_c$ and $\Theta_c$ as the parameters for the constant component $p_ {cons}$ and $\theta_{cons}$, respectively. To determine the ratio for the variable to constant component flux ($ I_{v/c} = I_ {var} /I_{cons}$), we used data from the night JD 2455685 where the contribution to the polarization of the variable component is almost zero, and the polarization degree is 1.5$\pm$0.8\%. In this way, we fitted the observed polarization finding that the ratio $I_{v/c}$= 1.79$\pm$0.24. This corresponds to a constant component flux value of $I_{c}$=1.47$\pm$0.19 mJy. With this result we solved the equations directly and found the values of $p_{var}$, $\theta_{var}$ and $I_{var}$ for each night. These results are presented for Cycles I and II in Table~\ref{tbl-4}, and are plotted in Figure~\ref{fig7}, where the polarimetric parameters of the variable component are shown during the low and high activity phases. In this figure it can be seen that during the low-activity phase (Cycle I) the polarimetric parameters of the constant polarized component, that are represented with a horizontal dashed-line, dominate over the variable polarized one (black-filled points). Therefore, for this cycle the observed polarization degree represented by empty squares, is much lower than the polarization degree of the variable component with a maximum difference of 42\%. Meanwhile, for the polarization angle the corresponding values for the variable and observed component, the difference is $\sim 8\%$. For Cycle II, we can see that the observed polarimetric values (polarization degree and angle) are very similar to the variable component. This result shows that the variable polarization component dominates over the constant component. Using data from Table~\ref{tbl-2}, columns (7) to (9), we find that the polarization degree is greater and more variable during the low-activity phase. 

From this analysis we can infer that the observed polarimetric behavior in the $R$-band can be interpreted as the superposition of two optically-thin synchrotron components, one stable and the other variable. The polarization degree $p_{cons}\sim 4\%$ derived for the stable component is identical to the value found by \citet{2004MNRAS.352..112B}, while the optical position angle $\theta_{cons}$  shows value that coincides with the observed $\theta \sim 150^\circ$. This value is similar to the projected position angle of the radio jet found by \citet{2008ApJ...678...64P,2010ApJ...723.1150P}. On the other hand, the source shows a larger polarization degree during the low-activity phase in comparison with the high-activity phase. The maximum polarization angle variations during the low-activity phase are $\sim$50$^{\circ}$. These variations are similar to the value found by  \citet{2003AJ....125.1060R} and \citet{2004MNRAS.352..112B} for the projected opening angle of the radio jet, while the maximum variations found during the high-activity phase are 10$\degr$ and 18$\degr$, for Cycle IIa and Cycle IIb, respectively.

The agreement between our optical results and those at radio bands (see above) suggests that there is a common magnetic field component for both the optical and radio bands, that is persistent in time and whose direction is transverse to the flow, as expected from a tangled magnetic field compressed by a shock.
The polarimetric characteristics found in this work are within the framework of a spine-sheath model proposed by \citet{2005A&A...432..401G} and used for this source by \citet{2010ApJ...723.1150P}. In this context, the stable component would be associated with the sheath and the variable component with the spine.

\subsection{Standing Shock Scenario}

A standing shock scenario has already been proposed by \citet{2008ApJ...679.1029T,2008ICRC....3.1021H} to explain the variable behavior of this object. A stationary shock wave system in a relativistic jet can be caused by the pressure imbalance between the jet and the surrounding ambient medium. Oscillations of the jet width occur as the overpressured medium in the jet expands until its pressure falls below the ambient pressure. This underpressured plasma then contracts under the influence of the external medium until high pressure is restored. This leads to the formation of a system of standing oblique shock waves, perhaps terminating in a strong shock perpendicular to the jet flow \citep[see][and references therein]{1997ApJ...482L..33G,2004ApJ...613..725S}. To identify the origin of the variations in flux and polarization, we must consider that the two flares observed in 2009 and 2010 had time-scales of $\sim$months. For this object, we have assumed that the radiation originates in the same region of the jet in a standing shock. 

Let us assume that the variable $R$-band flux component is produced by a shock. In the one-zone homogeneous SSC model, the source is assumed to be a spherical blob of radius $r_{b}$ moving in a turbulent plasma, with a constant jet's Lorentz factor $\Gamma_{j}$, such that the emission region size is considered stable. Thus, in the observer's reference frame, the flux of the shocked region is amplified as:
\begin{equation}
F=F_0\nu^{-\alpha}\delta^{(3+\alpha)}\delta'^{(2+\alpha)} \;\;,
\label{F0F}
\end{equation}
where $\delta=[\Gamma_{j}(1-\beta\cos\Phi]^{-1}$ is the jet's Doppler factor, $\beta = (1-\Gamma_{j}^{-2})^{1/2}$ its global velocity in units of speed of light, and $\Phi$ is the viewing angle. The factor 
$\delta'$ is the Doppler factor of the shocked plasma in the shock's front reference frame. To determine $\delta'$ it would be necessary to know the plasma's state equation, nevertheless it cannot be determined since most of its parameters are unknown.  However, without loss of generality it can be assumed that the plasma speed in the shock's reference frame should be much less than the
speed of light $(<< c)$ , assuming that  $\delta' \approx1$ \citep{2008ApJ...672...40H}.

The observed degree of polarization depends on the shock's viewing angle  $\Psi$, the spectral index $\alpha$, and the ratio of densities of the shocked region to the unshocked region $\eta = \eta_{shock}/\eta_{unshock}$ \citep{1991bja..book....1H}:

\begin{equation}
  p\approx\frac{\alpha+1}{\alpha+5/3}\frac{(1-\eta^{-2})\sin^2\Psi}{2-(1-\eta^{-2})\sin^2\Psi} \;\; ,
\label{F1F}
\end{equation}
and
\begin{equation}
  \Psi=\tan^{-1}\left\{\frac{\sin\Phi}{\Gamma_{j}(\cos\Phi-\sqrt{1-\Gamma_{j}^{-2}})}\right\}
  \;\; .
\label{F2F}
\end{equation}

Following \citet{2008ApJ...679.1029T}, we assumed a bulk Lorentz factor $\Gamma_j$ = 18 for the SED of 1ES 1959+650.
We also have used the value of $\alpha_{ox}=1.64$, given by \citet{2010ApJ...716...30A}. 

From equation ~(\ref{F0F}) we can estimate the Doppler factor as a function of time. The value of $F_0$ is determined by $F_0 = F_{max} \nu^{\alpha} /\delta_0^{(3+ \alpha)}$, where $F_{max}$ is the maximum observed flux and $\delta_0$ is obtained from $\Phi_0$, which is calculated from equations~(\ref{F1F}) and~(\ref{F2F}) for the maximum value of the polarization degree of the variable component  (see Table~\ref{tbl-3}). From \citet{1991bja..book....1H} we adopt $\eta=2.3$, and $\Psi = \pi/2$ which is the minimum possible compression that produces a degree of linear polarization as high as 47$\%$. This yields $\Psi_0=72.9\degr$, $\Phi_0= 2.35\degr$, and $\delta_0$=23.3 at the maximum polarization (see Table~\ref{tbl-5}).

Applying equation~(\ref{F0F}) to the photometric data, we estimate the Doppler factor $\delta(t)$ and thus, the viewing angle of the jet $\Phi(t)$ as functions of time. The latter allows us to estimate the viewing angle of the shock $\Psi(t)$ (eq. \ref{F2F}). Figure~\ref{fig8} shows the values of the Doppler factor, viewing angle of the jet and the viewing angle of the shock derived from the observed flux variability in Cycle II. Applying equation~(\ref{F1F}) to the polarimetric data, we obtain the value of the plasma compression as function of time.  The value of $\eta(t)$ is shown in Figure~\ref{fig8} as well. All these parameters are presented in Table~\ref{tbl-6}, where we give in column (1) the Julian Date and in columns (2) to (5) the values obtained for $\delta(t)$, $\Phi(t), \Psi(t)$ and $\eta(t)$ along with their errors, respectively. 

From Figure~\ref{fig8}, it can be seen that when the source shows its maximum brightness (14.08 mag, JD 2455002), the Doppler factor reaches 23.3, while in the minimum (14.74 mag, JD 2455150) it is 20.4. This corresponds to a maximum variation of $\Delta\delta$ = 2.9. The viewing angle of the jet $\Phi$ is anti-correlated with the flux, showing a minimum value of 2.35$\degr$ when there is a maximum in brightness, and a maximum value of  2.78$\degr$. Therefore, it shows a maximum variation of $\Delta \Phi$= 0.43$\degr$. These small variations can produce large flux variations while $\Gamma_j$ remains constant. In the state of maximum brightness, the viewing angle of the shock $\Psi\sim73\degr$ undergoes its maximum aberration due to relativistic effects, deviating $\Delta\Psi\sim17\degr$ with respect to the shock transverse plane, while in a state of minimum brightness the shock tends to align perpendicularly to the axis of the radio jet. This finding may explain the high degree of polarization found in the low activity phase. That is, the amplification of the magnetic field components parallel to the shock due to Doppler effect is not enough to considerably increase the polarization degree. In this context, the shock could be affecting an emission zone with a magnetic field almost parallel to the jet axis. Another possibility is that the polarization degree may depend more on other factors, for example, the degree of compression in the electron density or depolarization by other components. 
We find the maximum compression of the plasma $\eta$=1.145 when the polarization degree reaches its maximum value of $10.3\%$ (JD 2455062). We got the minimum compression factor $\eta_{min}$=1.052 when the polarization degree has a minimum value of 3.8$\%$ (JD 2454974). Thus, small changes in the compression factor $\Delta\eta \approx 0.093$ can produce large changes in the polarization degree.

\subsection{Magnetic Field Structure}
The lifetime of the synchrotron electrons for a given frequency $\nu$ in GHz \citep{2008ApJ...672...40H} 
is:

\begin{equation}
t_{var}=4.75\times10^2\left(\frac{1+z}{\delta_0\, \nu_{GHz}\,B^3}\right)^{1/2} \,\rm d \;\;.
\end{equation}

From our data we get $t_{var}\approx t_{min}=9.74\pm1.17$ d, and $\delta_0= 23.3$. From here, we obtain the magnetic field intensity $B=$~0.061$\pm$0.005~Gauss, and an upper limit to the emission region size of $r_{b}\leq ct_{min} \delta_0/(1+z)$ = (5.61$\pm$0.68)$\times10^{17}$ cm.

The behavior of the variable polarization strongly depends on the amount of ordering of the magnetic field inside the source. The polarimetric properties observed in inhomogeneous sources at
a given time and frequency will result from the integrated characteristics of all the different emitting regions. These inhomogeneities produce a depolarization degree in the emission source, revealing possible asymmetries in the magnetic field structure. In this work, we used the model proposed by \citet{1988ApJ...332..678J}, where the inhomogeneities are produced by hydromagnetic turbulence, breaking the coherence beyond a characteristic distance $l_B$. Inside the emitting region, the magnetic field is modeled as a net of cubic cells with size $l_B$,
which are subsequently divided into cells with size  $l_B/2$, until the observed results are fitted. Each $k$-cell contains magnetic field vector's $B_k$, so the field's structure is considered to be frozen inside the plasma as it moves through the jet. Except for the influence of shocks, the field turbulence is isotropic in the fluid frame. The total magnetic field is the sum of all the field vectors inside the net. In the shock, both the density and the field lines are compressed in such a way that the magnetic field components parallel to the shock's plane are amplified by relativistic beaming, whereas the normal components are less influenced by relativistic effects \citep[see][]{1985ApJ...298..301H}.

Most of our data show that there is a lack of correlation between the total flux and the polarization degree. This result suggests that the evolution of the magnetic field is decoupled from the acceleration of the particles produced by the shock. An alternative way to obtain the structure of the magnetic field is through the calculation of the coherence length of the magnetic field at large scale $l_B=(\kappa
\Pi_0/\sigma_p)^{-2/3} l$, where $\Pi_0 = 0.75$ is the fraction of polarization in a region with magnetic field perfectly ordered, $\kappa$ the degree of intrinsic polarization of the source, $\sigma_p$ the constant polarization dispersion, and $l$ the emission region size \citep[see][]{1985ApJ...290..627J,1988ApJ...332..678J}. 
Assuming that $\kappa$ is the maximum value of polarization for the variable component ($\kappa$=47\%), taking $\sigma_p=2$ \% and $l \sim r_b$, we obtain $ l_B=1.31\pm0.16\times10^{17}$ cm.

Observational and theoretical studies in radio \citep{1985ApJ...298..301H,1985ApJ...298..114M}, show that variations in polarization on time-scales of several months are directly related to the distance along the jet traveled by the relativistic shock in the time between two extremes of the polarization light curve. If $r_s$ is this distance, then the observed time scale $\Delta t$ is given by \citep{1991A&A...241...15Q}
\begin{equation}
\Delta t=\frac{r_s(1+z)}{c\beta_s\delta_s \Gamma_s}\;\;,
\end{equation}
where $c\beta_s$ is the speed of the shock, $\delta_s$ and $\Gamma_s$ are the Doppler factor and Lorentz factor of the shock, respectively. Using values derived in radio at 43 GHz by \citet{2004ApJ...600..115P,2008ApJ...678...64P,2010ApJ...723.1150P} for the shock speed ($\beta_s= 0.1$), Lorentz factor of the shock ($\Gamma_s\approx 3$), and Doppler factor of the shock ($\delta_s\approx 2\Gamma_s$), and taking the minimum observed time scale between two extremes of the polarization light curve, $\Delta t$=29d (Cycle III), we obtain $r_s \approx$ (1.29 $\pm$ 0.26)$\times 10^{17}$~cm for the distance traveled by the shock. This distance is consistent with the coherence length  $l_B$, suggesting a connection between the monthly variations observed in the polarization degree and the spatial changes in the magnetic field induced by the inhomogeneities in the jet. All these derived parameters are summarized in Table~\ref{tbl-5}.

From our polarimetric analysis we find that our results are consistent with the results found at radio bands by \citet{2010ApJ...723.1150P}. This work suggests that the behavior of the polarimetric parameters in radio are in good agreement with the spine-sheath model when applied to 1ES1959+650. In this context, the electric vector position angle (EVPA) shows two different positions, one parallel along the jet axis, and another orthogonal at the edges of the jet. These authors also find that the polarization degree decreases as the jet is aligned toward the observer and increases at the edges of the jet. 

\section{SUMMARY}
\label{Conc}

During the optical polarimetric observations of 1ES~1959+650 the object showed in the $R$-band light curves mainly a period of low activity phase (2008), and later two active phases (flares) in 2009 and 2010. The time elapsed between the two maxima (14.08 mag at JD 2455002 and 14.28 mag at JD 2455505) was $\sim$500 d. From a detailed photometric and polarimetric analysis we have found the following results:

1. The minimum variability timescale of the $R$-band flux was found to be $\sim$10 d for the entire period of observations. This result is similar to the value found by \citet{2004ApJ...601..151K} and \citet{2008ApJ...679.1029T}. We also found a maximum brightness variation of
$\Delta m_{R}$=1.12 mag in $\Delta t$=329 d. During the first maximum ($R$=14.08 mag) the brightness increases at a rate of 0.026 mJy/d and decreases at a rate of 0.025 mJy/d. During the second maximum ($R$=14.28 mag) the source displayed a steady brightness increase from minimum brightness of $R$=15.2 mag to maximum brightness of $R$=14.28 mag, with a rate of 0.020 mJy/d. In the entire period of observations, the source showed a minimum and maximum brightness of $R$=15.2 mag and $R$=14.08 mag, respectively.

\begin{figure}
\epsscale{1.0}
\plotone{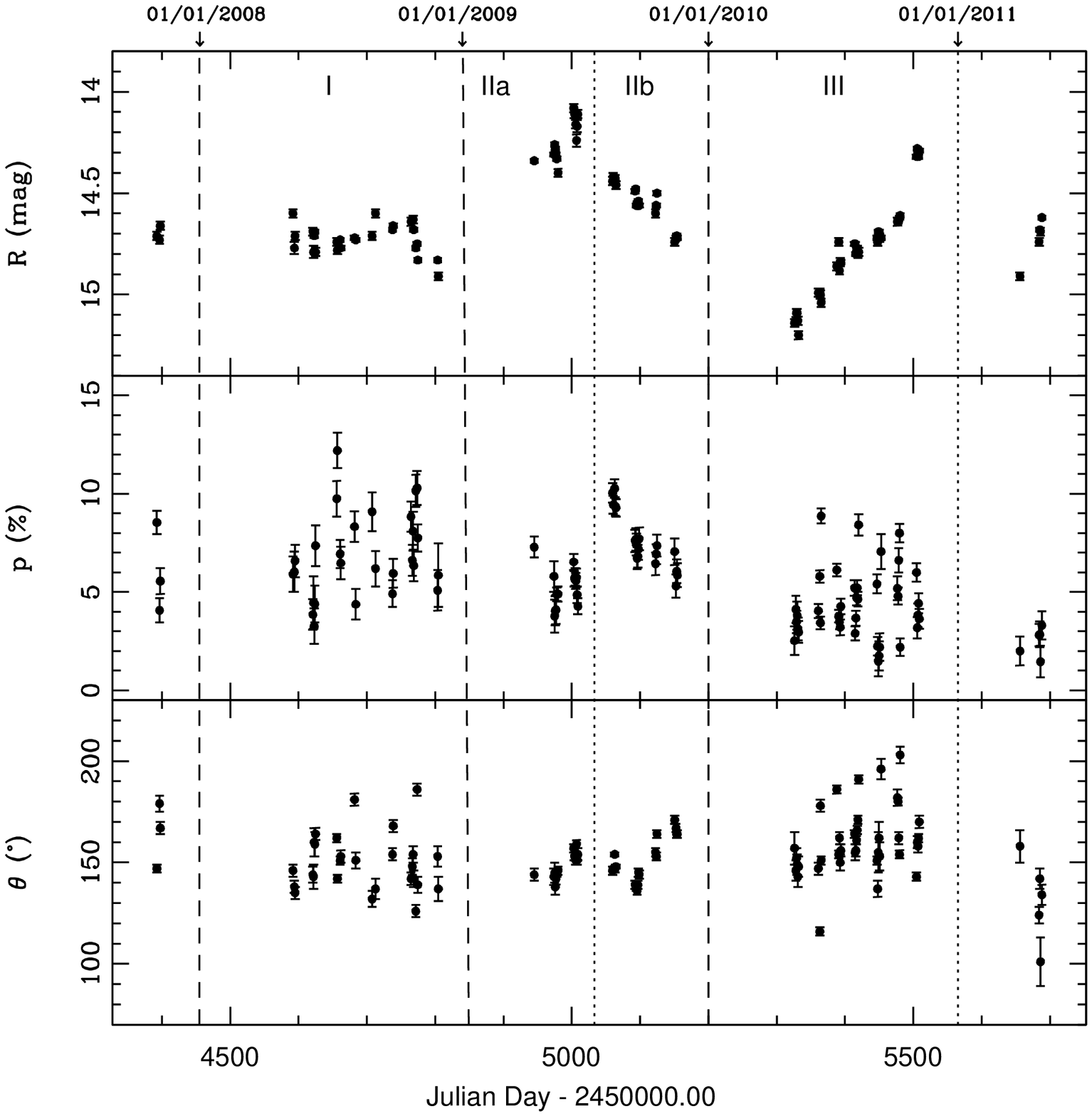}
\caption{Light curves showing that the object displayed variations with timescales of $\sim$days-months. In top panel, the $R$-band (mag);
middle panel, the polarization degree P(\%); and bottom panel, the polarization position angle
$\theta$($^{\circ}$) vs Julian Date. Data are from 2007 October 18 to 2011 May 05. Each point has an 
associated error bar and corrected by the contribution of the host-galaxy, see text.\label{fig1}}
\end{figure}

\begin{figure}
%\epsscale{0.5}
\plotone{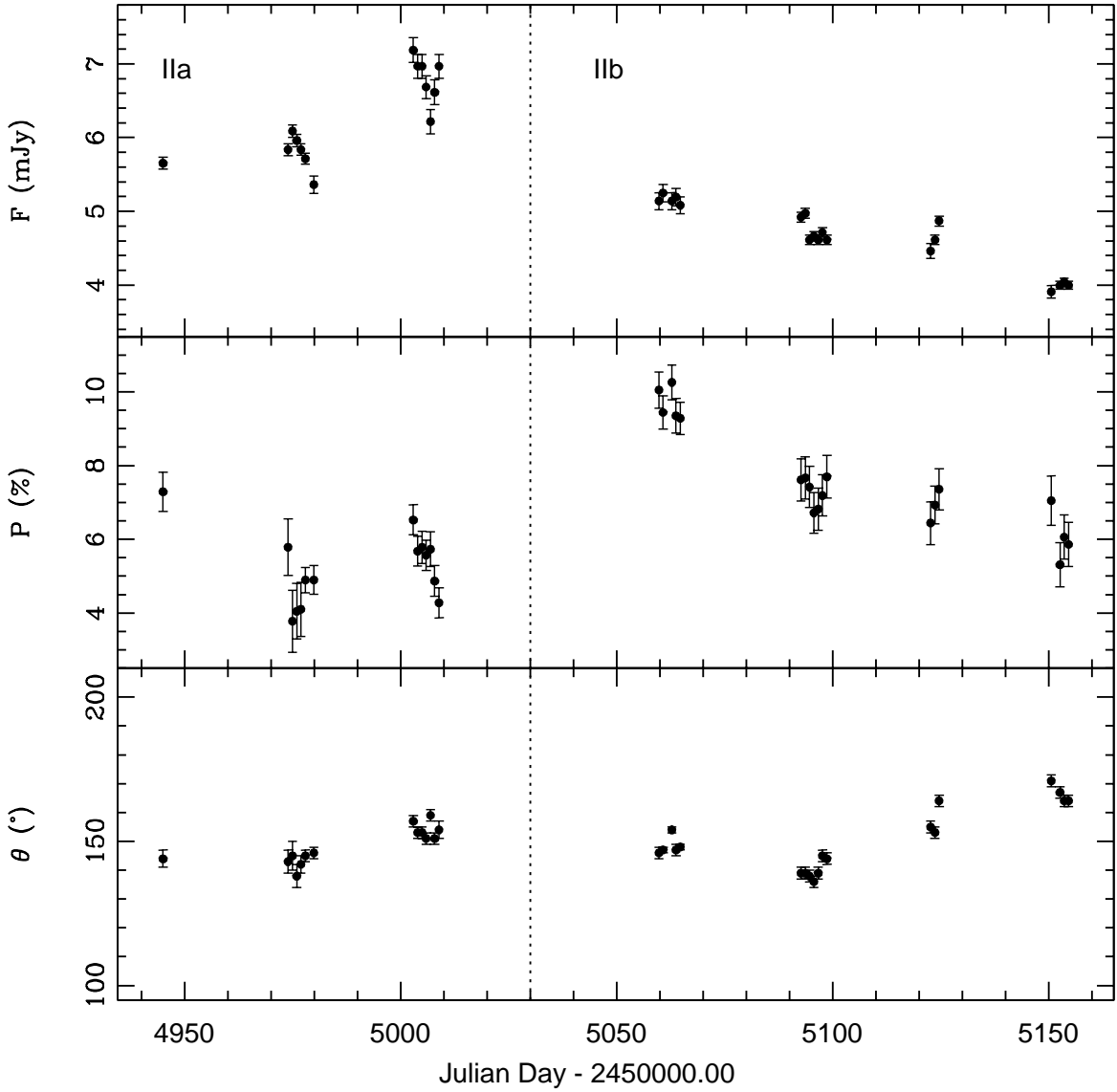}
\caption{A blow-up of the polarimetric parameter shown in Figure~\ref{fig1} for
Cycles IIa and IIb. There can be appreciated a clear correlation between the
$R$-band flux and the polarization degree during Cycle IIb, while the polarization
angle varied around the average value of $\sim$ 150$^\circ$.\label{fig2}}
\end{figure}

\begin{figure}
%\epsscale{0.5}
\plotone{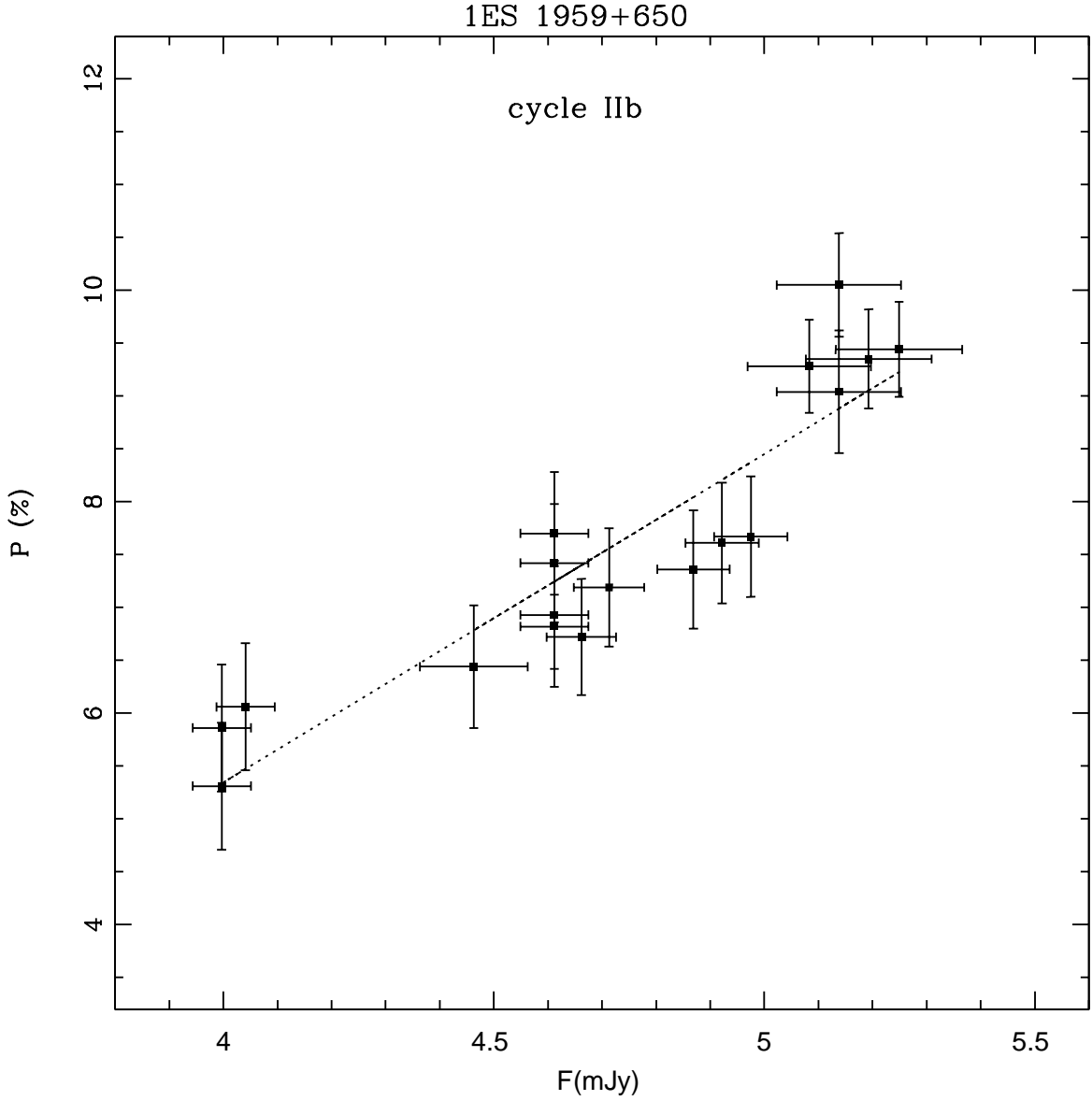}
\caption{Correlation between the $R$-band flux and polarization degree for 
Cycle IIb (2009 August 14 to November 18). The Pearson's correlation
coefficient is  $r_{pol}$=0.984$\pm$0.025. There is a positive 
correlation between both parameters.\label{fig3}}
\end{figure}

\begin{figure}
\epsscale{1.0}
\plotone{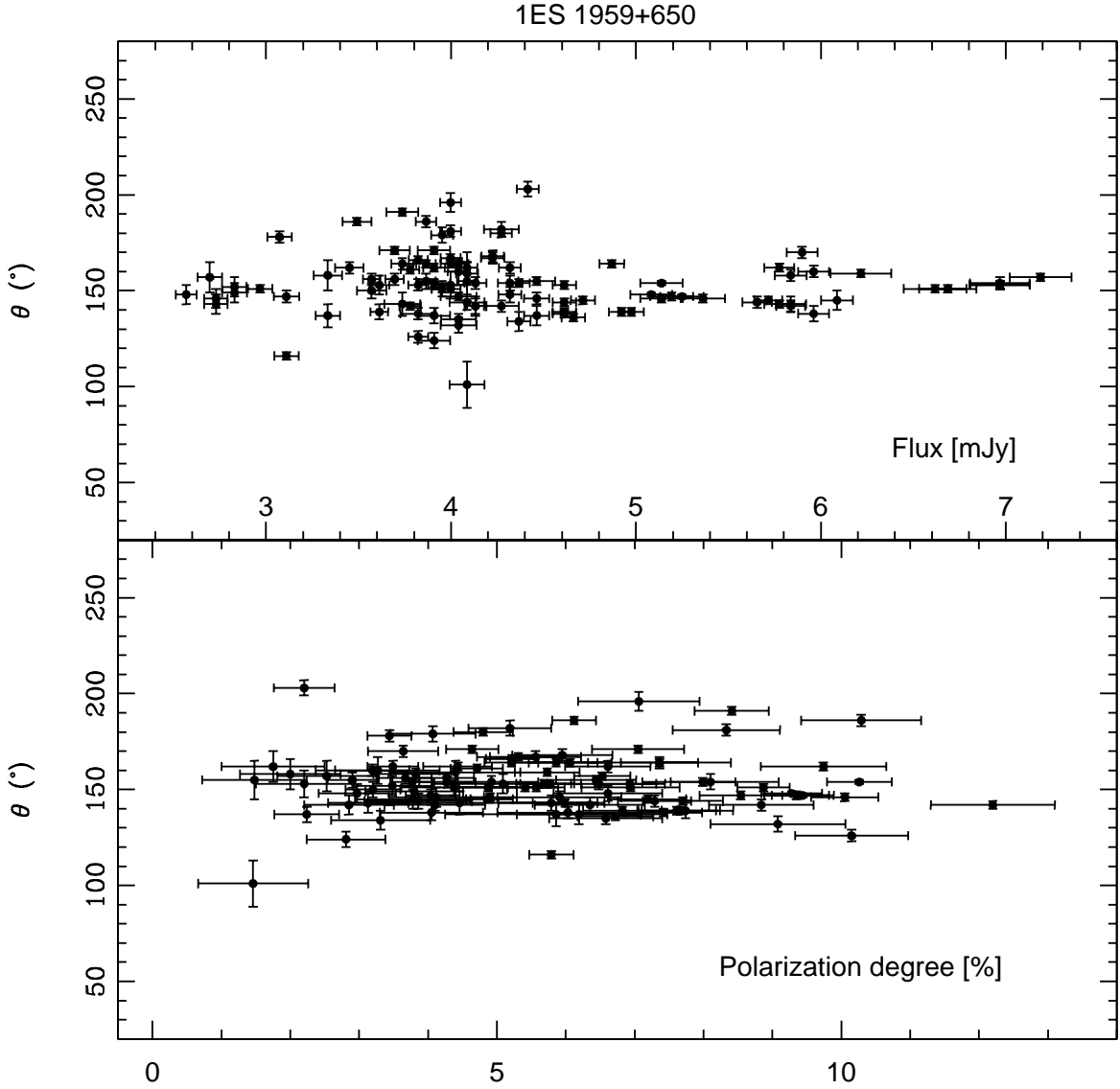}
\caption{Top Panel: Correlation between the polarization angle and the $R$-band flux. Bottom panel:
Correlation between the polarization angle and the polarization degree. In both panels there can be noticed a preferential tendency of the position angle $\sim\,150^{\circ}$.\label{fig4}}
\end{figure}

\begin{figure}
%\epsscale{0.5}
\plotone{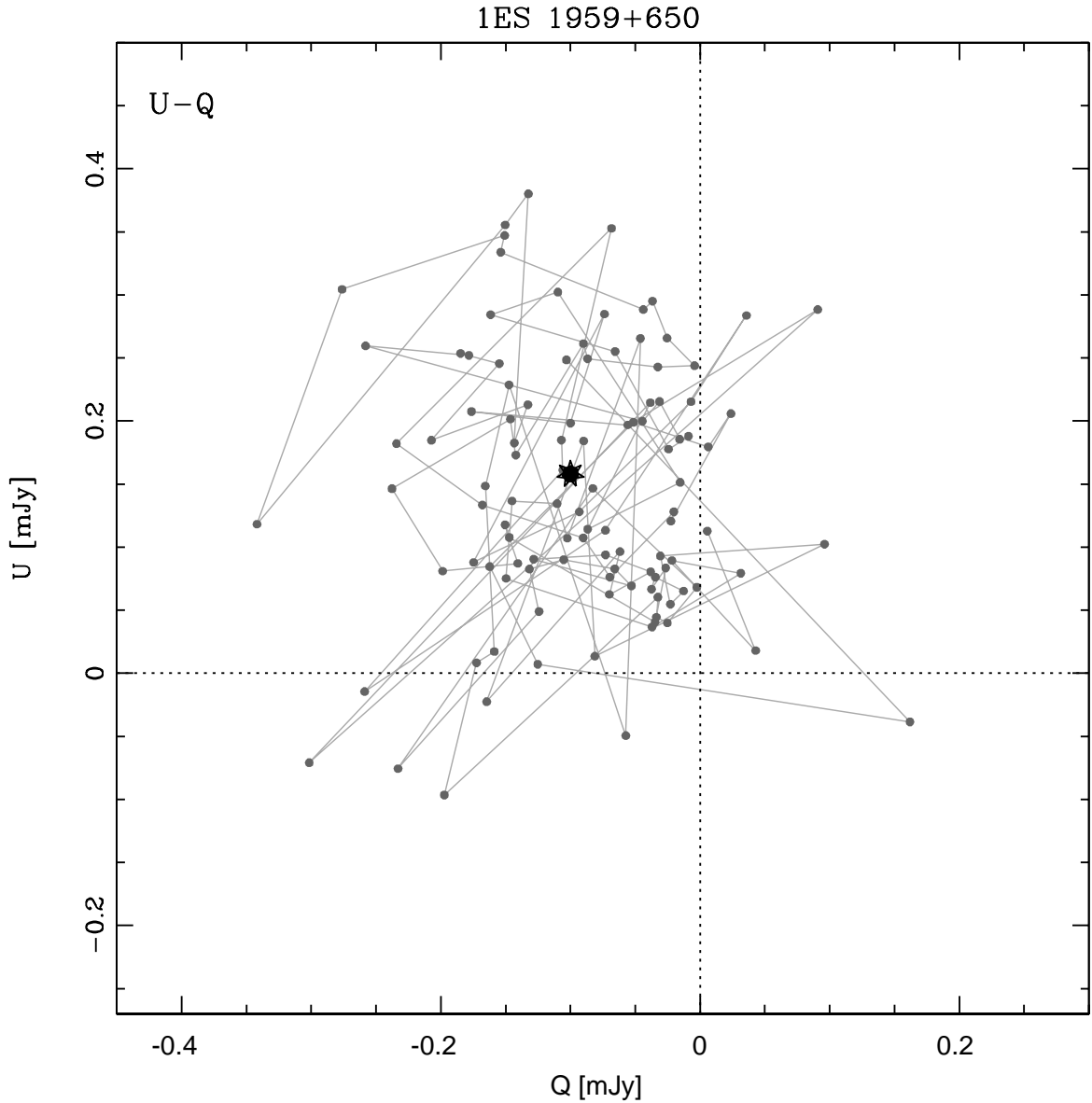}
\caption{Q-U absolute Stokes parameters plane obtained using
all data. The solid black-filled star marks the mean constant value and the
existence of a stable polarization component. The obtained
average values are: $\cal h$Q$\cal i$= - 0.10$\pm$0.01 mJy and $\cal
h$U$\cal i$ = 0.16$\pm$0.02 mJy. The polarization
degree and position angle values obtained for the constant component are
$P_c=(4.1\pm0.5)\%$ and $\Theta_c=151\degr\pm13\degr$, 
respectively. The constant polarization degree dispersion is
$\sigma_p$=2.1$\%$.\label{fig5}}
\end{figure}

\begin{figure}
%\epsscale{0.5}
\plottwo{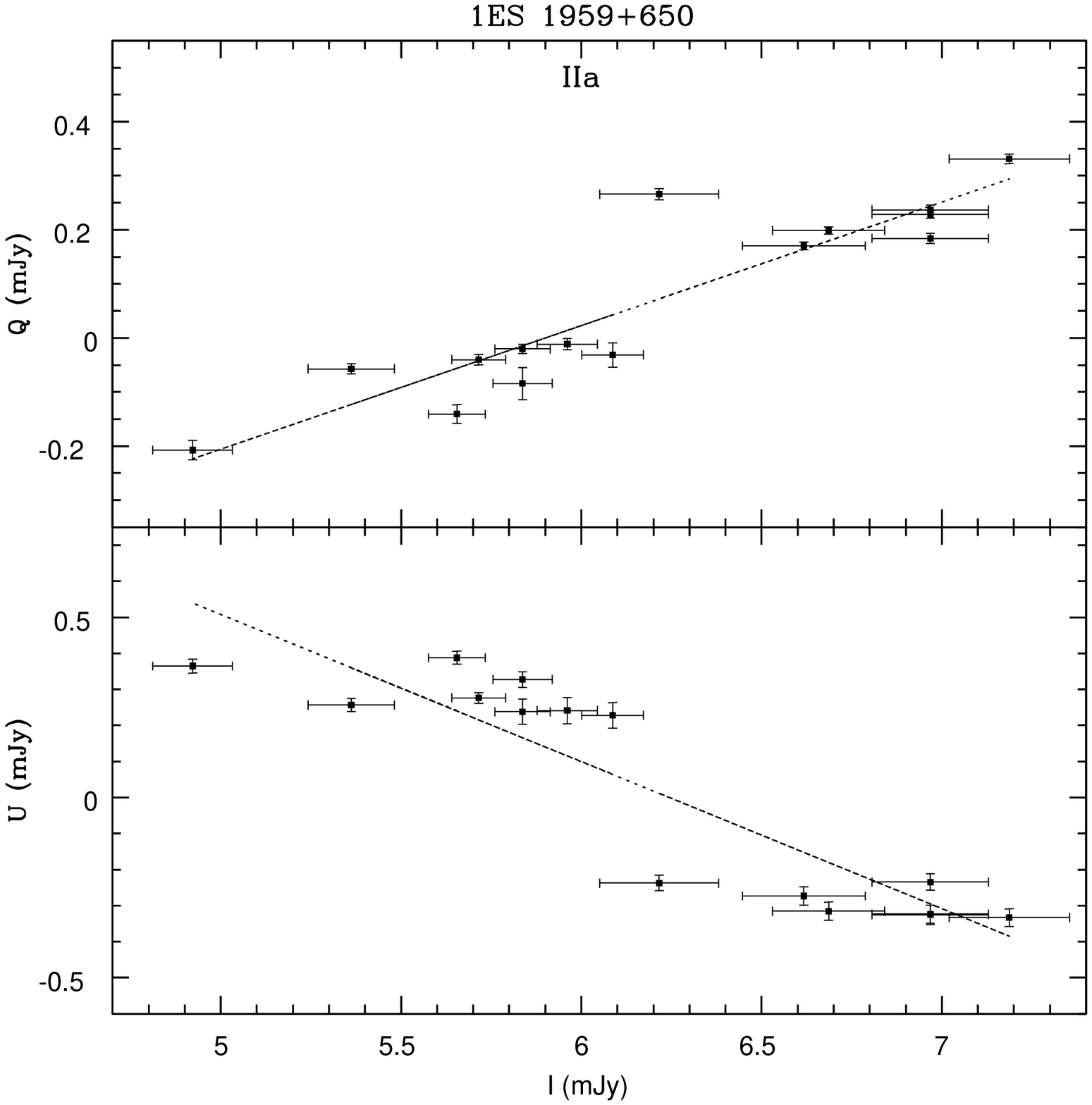}{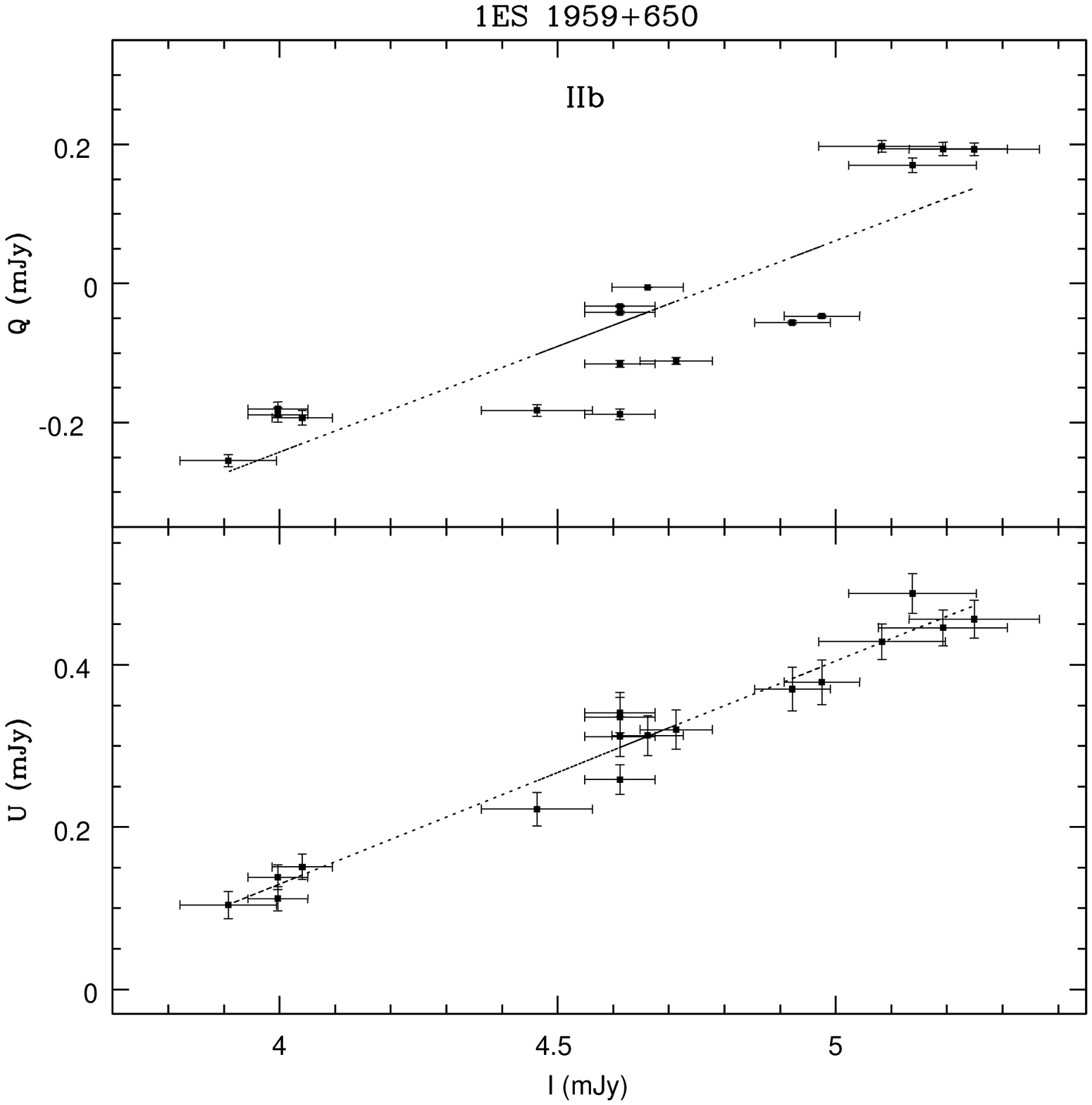}
\caption{  Left: linear correlation between the Stokes parameters Q
vs I (top panel), and U vs I (bottom panel) for Cycle IIa (2009 Apr-Jun). The correlation coefficients obtained for these parameters 
are  $r_q$=0.913 and $r_u$= -0.902, and the slopes $m_q$=0.228$\pm$0.028,
$m_u$=-0.408$\pm$0.054, respectively. Right: linear correlation between Q vs I (top panel), and U vs I (bottom panel) for Cycle IIb. The correlation coefficients for this Cycle IIb are
$r_q$=0.872 and $r_u$=0.980, and slopes  $m_q$=0.304$\pm$0.044,
$m_u$=0.275$\pm$0.015, respectively. 
\label{fig6}}
\end{figure}

\begin{figure}
\plottwo{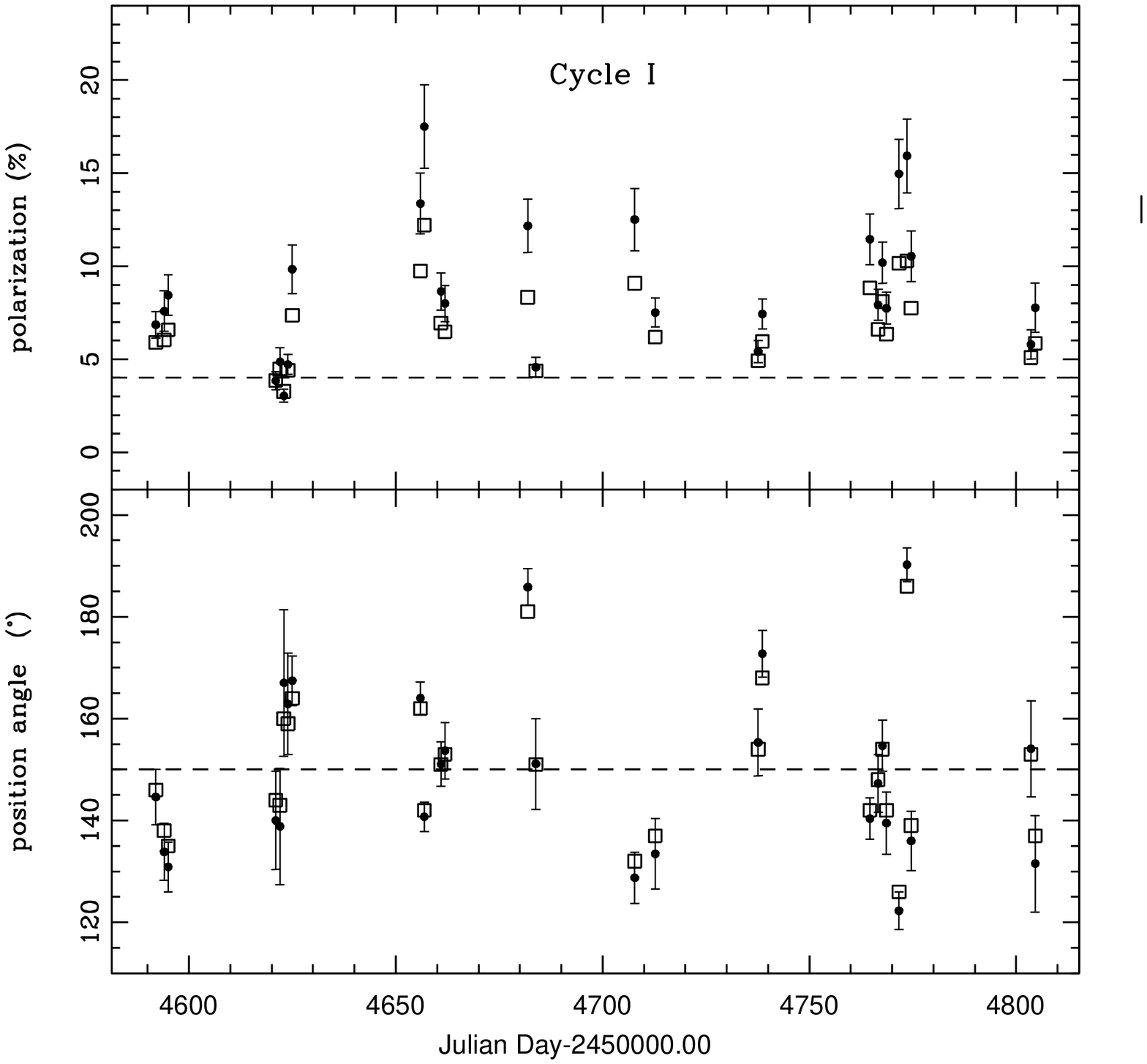}{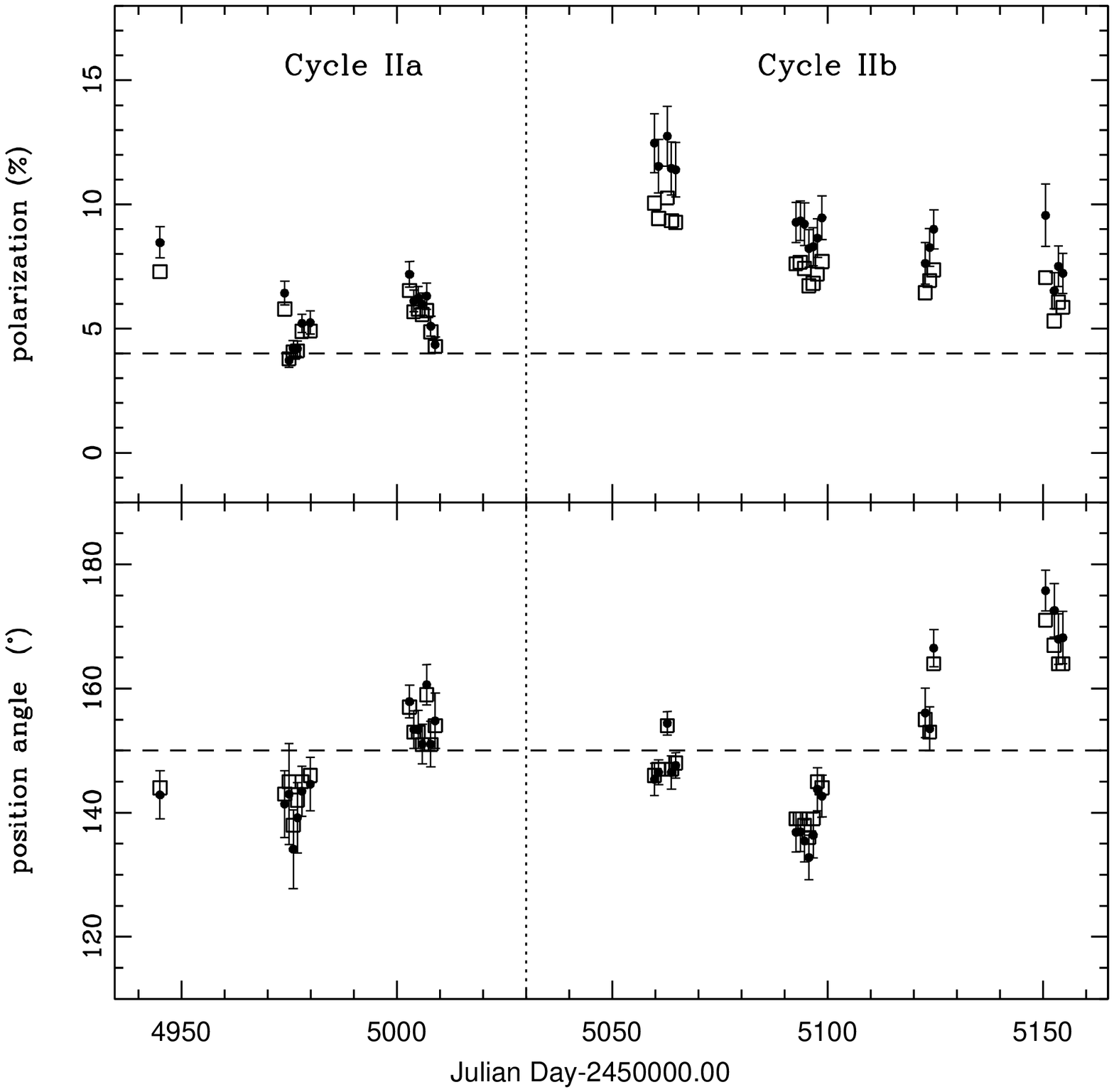}
\caption{Data collected during the low activity state in 2008 (left) and later during the high activity state in 2009 (right).
They were fitted using a two-component synchrotron model. Squares show the superposition of a constant
(dashed lines) and a variable polarized component (solid dots) using the addition law for polarized sources. 
In both figures, top panel shows the polarization degree and bottom panel shows the polarization
angle. The model values are presented in Table~4. The values for the constant component presented in the plots are $p_{cons}=(4.1\pm0.5)\%$ and $\theta_{cons}=151\degr\pm13\degr$. During the high activity state in 2009 (right), the variable component
dominates over the constant component. \label{fig7}}
\end{figure}

\begin{figure}
\plotone{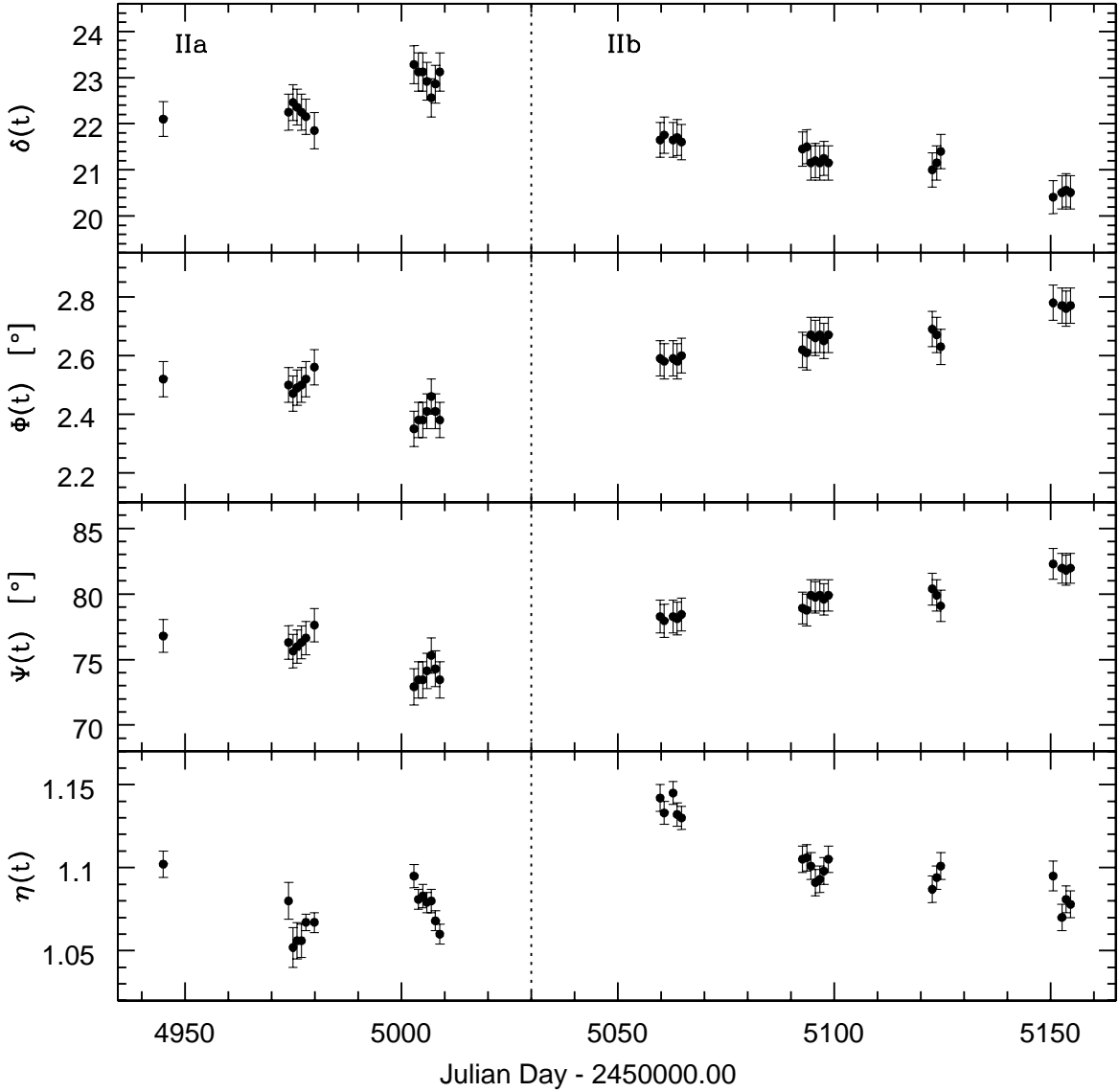}
\caption{Temporal variability of some physical parameters related to the relativistic jet kinematics of 1ES 1959+650 derived in this work.
From top to bottom panels: the Doppler factor, $\delta(t)$; the angle between the
jet axis and the line of sight, $\Phi(t)$;  viewing angle of the shock, $\Psi(t)$; and the compression factor of the shocked
to the unshocked plasma, $\eta(t)$, during the high activity state in 2009. \label{fig8}}
\end{figure}

2. In general, there could not be found any correlation between the polarization degree and the optical flux, except when the source started to decrease its brightness after the first maximum occurred. This can be seen in Figure~\ref{fig1} during Cycle IIb (2009 August 18, JD 2455062 to November 16, JD 2455152). Therefore, only for Cycle IIb a correlation between the optical R-band flux and the degree of linear polarization was found, with coefficient of correlation $r_{pol}=0.984\pm0.025$. The maximum polarization degree of 12.2\% was observed during the low activity phase, and the maximum variation of 10.7\% occurred on a timescale of 2.8 yr. It is worth to mention that the polarization degree shows a more variable behavior  ($\mu\sim 32\%$) with a larger amplitude ($Y\sim 128\%$) during the  low-activity phase than during the higher activity phases where $\mu \sim 19\%$ and $Y\sim 64\%$. The minimum polarization degree time-scale is $\sim$29 d. Furthermore, the source showed a larger average polarization degree during the low-activity phase (6.9\%) in comparison with the high-activity phase (5.2\%).
No polarization angle rotation is observed during the monitored period of time. Instead, the source presented a preferential average position angle of 153$\pm$16$\degr$. This preferred position angle of the optical polarization is in good agreement with the projected position angle of the parsec scale jet found by \citet{2010ApJ...723.1150P} from 43 GHz images. The lager dispersion of $\sim$50$\degr$ for the polarization angle was found during the low-activity phase. This value is similar to the projected opening angle of the jet reported by \citet{2003AJ....125.1060R}. In the high-activity phase the dispersion turned out to be lower, i.e. $\sim$18$\degr$.

3. From the analysis done with the Stokes parameters $\{Q,U,I\}$, there follows the existence of two components that contribute to the polarized flux, one stable and another variable. The stable component has a constant polarization degree $p_{cons}=4\%$  and a polarization angle 
$\theta_{cons}\sim150\degr$. These results are consistent with radio interferometric observations done by \citet{2004MNRAS.352..112B}. In this context, the polarimetric behavior is consistent with a spine-sheath structure of the jet where the variable component can be associated with the central jet structure taking the form of spine and the constant component with a stable structure in the form of sheath surrounding the jet.

4. The $R$-band observed variations can be explained in the framework of a standing-shock model. Using the spectral index  $\alpha_{ox}$ = 1.64 given by \citet{2010ApJ...716...30A}, and a bulk Lorentz factor of the jet $\Gamma_j=18$  found by \citet{2008ApJ...679.1029T}, a Doppler factor $\delta_0 =$ 23.3, and a viewing angle of the jet $\Phi_0 = 2.35\degr$ were estimated. 
From our minimum variability timescale $t_{min}$ we obtained the magnetic field intensity associated to the jet  $B=0.06$ Gauss, and the size of the emission region  $r_b=5.6\times10^{17}$ cm. The magnetic field intensity obtained in this work is in agreement with the value reported by  \citet{2004ApJ...601..151K}, although the size of the emission region estimated by these authors ($r_b\approx 1.4 \times 10^{16}$ cm) is lower by an order of magnitude. 

5. Temporal variations of some parameters of the relativistic jet of 1ES 1959+650 were derived:
(a) using photometric data, the Doppler factor $\delta (t)$, the viewing angle of the shock $\Psi(t)$, and the viewing angle of the jet $\Phi(t)$; (b) using polarimetric data, the ratio between the shocked to unshocked densities $\eta(t)$ and the coherence length of the magnetic field $l_B$=1.31$\times10^{17}$ cm. 
  
6. From our data, and using values derived by \citet{2004ApJ...600..115P,2008ApJ...678...64P,2010ApJ...723.1150P} for the physical parameters of the shock $\beta_s$ and  $\Gamma_s$, we obtained the distance traveled by the shock $r_s$= 1.29$\times10^{17}$ cm. This distance is consistent with the field turbulence scale $l_B$, suggesting a connection between variations of the polarization degree and the spatial changes in the magnetic field.

In a future work, an optical color polarization variability study will be carried out with the aim of studying the behavior of polarization with frequency. This will enable us to establish whether changes in the polarization degree found in this work are due to inhomogeneities in the density of particles in the emission zone or to changes in the magnetic field direction in the jet.

\acknowledgments
M.S. thanks CONACyT through grant 177304 for a graduate student fellowship. M.S., E.B., D.H., J.I.C., R.M. and J.H. acknowledge financial support from UNAM-DGAPA-PAPIIT through grant IN116211. I.A. acknowledges funding support by MINECO grant AYA2010-14844, and CEIC (Andaluc\'ia) grant P09-FQM-4784.
We all want to thank the OAN-SPM staff for the support given to this project.  In particular, we acknowledge Jorge Vald\'ez, Fernando Quir\'os, 
Benjam\'{\i}n Garc\'{\i}a, Enrique Colorado and Esteban Luna for their contributions to the design and construction of Polima. This research has made use of the SAO/NASA’s Astrophysics Data System (ADS) and of the NASA/IPAC Extragalactic Database (NED), which is operated by the Jet Propulsion Laboratory, California Institute of Technology, under contract with the National Aeronautics and Space Administration.

\bibliography{Sorcia.bib}

\clearpage

\clearpage

\clearpage

\clearpage

\clearpage

\clearpage

\clearpage

\clearpage

\clearpage

\begin{deluxetable}{lrrrrcrrrrr}
\tablewidth{0pt} \tablecaption{ POLARIZATION AND PHOTOMETRY IN THE $R$-BAND FOR 1ES 1959+650 \label{tbl-1}}
\tablehead{ 
\colhead{ }& \colhead{JD}& \colhead{$p$}& \colhead{$\epsilon_p$}&
\colhead{$\theta$}& \colhead{$\epsilon_{\theta}$}& \colhead{$R$}&
\colhead{$\epsilon_R$}& \colhead{$flux$}& \colhead{$\epsilon_{flux}$}\\
\colhead{ }& \colhead{2450000.00+}& \colhead{($\%$)}& \colhead{($\%$)}&
\colhead{($^\circ$)}& \colhead{($^\circ$)}& \colhead{(mag)}&
\colhead{(mag)}& \colhead{(mJy)}& \colhead{(mJy)}&
}
 \startdata  
&4392.6685 &   8.5 &  0.6 &  147 &  02 &  14.71 &  0.02 &  4.04 & 0.10 \\ 
&4396.6392 &   4.1 &  0.6 &  179 &  04 &  14.73 &  0.02 &  3.95 & 0.06 \\ 
&4397.6553 &   5.6 &  0.7 &  167 &  03 &  14.66 &  0.02 &  4.22 & 0.06 \\ 
\hline
Cycle I&4591.9907 &   5.9 &  0.9 &  146 &  03 &  14.60 &  0.02 &  4.46 & 0.07 \\ 
&4593.9912 &   6.0 &  1.0 &  138 &  03 &  14.77 &  0.03 &  3.82 & 0.10 \\ 
&4594.9883 &   6.6 &  0.8 &  135 &  03 &  14.71 &  0.02 &  4.04 & 0.09 \\ 
&4620.9575 &   3.9 &  0.8 &  144 &  04 &  14.69 &  0.02 &  4.09 & 0.09 \\ 
&4621.9854 &   4.5 &  1.3 &  143 &  06 &  14.79 &  0.03 &  3.74 & 0.10 \\ 
&4622.9287 &   3.3 &  0.9 &  160 &  07 &  14.71 &  0.01 &  4.04 & 0.06 \\ 
&4623.9331 &   4.4 &  0.9 &  159 &  06 &  14.69 &  0.01 &  4.09 & 0.06 \\ 
&4624.9839 &   7.4 &  1.0 &  164 &  03 &  14.79 &  0.02 &  3.74 & 0.06 \\ 
&4655.9331 &   9.7 &  0.9 &  162 &  02 &  14.74 &  0.02 &  3.91 & 0.06 \\ 
&4656.9160 &  12.2 &  0.9 &  142 &  02 &  14.78 &  0.02 &  3.78 & 0.06 \\ 
&4660.9390 &   6.9 &  0.7 &  151 &  02 &  14.73 &  0.01 &  3.95 & 0.06 \\ 
&4661.8857 &   6.5 &  0.8 &  153 &  03 &  14.77 &  0.01 &  3.82 & 0.05 \\ 
&4681.9053 &   8.3 &  0.8 &  181 &  03 &  14.72 &  0.01 &  4.00 & 0.06 \\ 
&4683.8711 &   4.4 &  0.8 &  151 &  04 &  14.73 &  0.01 &  3.95 & 0.06 \\ 
&4707.7686 &   9.1 &  1.0 &  132 &  04 &  14.71 &  0.02 &  4.04 & 0.10 \\ 
&4712.7568 &   6.2 &  0.9 &  137 &  05 &  14.60 &  0.02 &  4.46 & 0.07 \\ 
&4737.6260 &   4.9 &  0.7 &  154 &  03 &  14.68 &  0.01 &  4.13 & 0.06 \\ 
&4738.6494 &   5.9 &  0.7 &  168 &  03 &  14.66 &  0.01 &  4.22 & 0.06 \\ 
&4764.6245 &   8.8 &  0.8 &  142 &  03 &  14.64 &  0.02 &  4.27 & 0.09 \\ 
&4766.6401 &   6.6 &  0.8 &  148 &  04 &  14.63 &  0.01 &  4.32 & 0.06 \\ 
&4767.6763 &   8.1 &  1.0 &  154 &  04 &  14.63 &  0.02 &  4.32 & 0.07 \\ 
&4768.6421 &   6.4 &  0.8 &  142 &  04 &  14.68 &  0.01 &  4.13 & 0.06 \\ 
&4771.6611 &  10.1 &  0.8 &  126 &  03 &  14.77 &  0.01 &  3.82 & 0.05 \\ 
&4773.6025 &  10.3 &  0.9 &  186 &  03 &  14.75 &  0.01 &  3.86 & 0.06 \\ 
&4774.6011 &   7.7 &  0.7 &  139 &  04 &  14.83 &  0.01 &  3.61 & 0.05 \\ 
&4803.6006 &   5.1 &  1.0 &  153 &  05 &  14.83 &  0.01 &  3.61 & 0.05 \\ 
&4804.5962 &   5.9 &  1.6 &  137 &  06 &  14.91 &  0.02 &  3.33 & 0.06 \\
\hline 
Cycle IIa&4944.9717 &   7.3 &  0.5 &  144 &  03 &  14.34 &  0.01 &  5.65 & 0.08 \\ 
&4973.9253 &   5.8 &  0.8 &  143 &  04 &  14.31 &  0.01 &  5.84 & 0.08 \\ 
&4974.9087 &   3.8 &  0.8 &  145 &  05 &  14.26 &  0.01 &  6.09 & 0.08 \\ 
&4975.9033 &   4.1 &  0.7 &  138 &  04 &  14.28 &  0.01 &  5.96 & 0.08 \\ 
&4976.8965 &   4.1 &  0.7 &  142 &  03 &  14.31 &  0.01 &  5.84 & 0.08 \\ 
&4977.9121 &   4.9 &  0.3 &  145 &  02 &  14.33 &  0.01 &  5.71 & 0.07 \\ 
&4979.8735 &   4.9 &  0.4 &  146 &  02 &  14.40 &  0.02 &  5.36 & 0.12 \\ 
&5002.9106 &   6.5 &  0.4 &  157 &  02 &  14.08 &  0.02 &  7.19 & 0.17 \\ 
&5003.8867 &   5.7 &  0.4 &  153 &  02 &  14.11 &  0.02 &  6.97 & 0.16 \\ 
&5004.9136 &   5.8 &  0.4 &  153 &  02 &  14.11 &  0.02 &  6.97 & 0.16 \\ 
&5005.8696 &   5.6 &  0.4 &  151 &  02 &  14.16 &  0.02 &  6.69 & 0.16 \\ 
&5006.8984 &   5.7 &  0.5 &  159 &  02 &  14.24 &  0.03 &  6.22 & 0.17 \\ 
&5007.8599 &   4.9 &  0.4 &  151 &  02 &  14.17 &  0.03 &  6.62 & 0.17 \\ 
&5008.8726 &   4.3 &  0.4 &  154 &  03 &  14.11 &  0.02 &  6.97 & 0.16 \\ 
\hline
Cycle IIb&5059.7646 &  10.1 &  0.5 &  146 &  02 &  14.44 &  0.02 &  5.14 & 0.12 \\ 
&5060.7266 &   9.4 &  0.5 &  147 &  01 &  14.42 &  0.02 &  5.25 & 0.12 \\ 
&5062.7378 &  10.3 &  0.5 &  154 &  01 &  14.44 &  0.02 &  5.14 & 0.12 \\ 
&5063.7241 &   9.4 &  0.5 &  147 &  02 &  14.43 &  0.02 &  5.19 & 0.12 \\ 
&5064.7183 &   9.3 &  0.4 &  148 &  01 &  14.46 &  0.02 &  5.08 & 0.11 \\ 
&5092.6494 &   7.6 &  0.6 &  139 &  02 &  14.49 &  0.01 &  4.92 & 0.07 \\ 
&5093.6284 &   7.7 &  0.6 &  139 &  02 &  14.48 &  0.01 &  4.98 & 0.07 \\ 
&5094.6235 &   7.4 &  0.6 &  138 &  02 &  14.56 &  0.01 &  4.61 & 0.06 \\ 
&5095.6250 &   6.7 &  0.6 &  136 &  02 &  14.55 &  0.01 &  4.66 & 0.06 \\ 
&5096.6221 &   6.8 &  0.6 &  139 &  02 &  14.56 &  0.01 &  4.61 & 0.06 \\ 
&5097.6221 &   7.2 &  0.6 &  145 &  02 &  14.54 &  0.01 &  4.71 & 0.06 \\ 
&5098.6680 &   7.7 &  0.6 &  144 &  02 &  14.56 &  0.01 &  4.61 & 0.06 \\ 
&5122.6343 &   6.4 &  0.6 &  155 &  02 &  14.60 &  0.02 &  4.46 & 0.10 \\ 
&5123.6606 &   6.9 &  0.5 &  153 &  02 &  14.56 &  0.01 &  4.61 & 0.06 \\ 
&5124.6235 &   7.4 &  0.6 &  164 &  02 &  14.50 &  0.01 &  4.87 & 0.07 \\ 
&5150.6030 &   7.0 &  0.7 &  171 &  02 &  14.74 &  0.02 &  3.91 & 0.09 \\ 
&5152.5942 &   5.3 &  0.6 &  167 &  02 &  14.72 &  0.01 &  4.00 & 0.05 \\ 
&5153.5879 &   6.1 &  0.6 &  164 &  02 &  14.71 &  0.01 &  4.04 & 0.05 \\ 
&5154.5879 &   5.9 &  0.6 &  164 &  02 &  14.72 &  0.01 &  4.00 & 0.05 \\
\hline 
Cycle III&5325.9937 &   2.5 &  0.7 &  157 &  08 &  15.14 &  0.02 &  2.70 & 0.06 \\ 
&5327.9604 &   4.1 &  0.7 &  146 &  05 &  15.13 &  0.02 &  2.73 & 0.06 \\ 
&5328.9199 &   3.5 &  0.6 &  152 &  05 &  15.09 &  0.02 &  2.83 & 0.07 \\ 
&5329.9141 &   3.8 &  0.7 &  149 &  05 &  15.09 &  0.02 &  2.83 & 0.07 \\ 
&5330.9302 &   3.1 &  0.6 &  143 &  05 &  15.13 &  0.02 &  2.73 & 0.06 \\ 
&5331.9224 &   3.0 &  0.6 &  148 &  05 &  15.20 &  0.02 &  2.57 & 0.06 \\ 
&5360.8804 &   4.0 &  0.4 &  147 &  03 &  14.99 &  0.02 &  3.11 & 0.07 \\ 
&5362.9268 &   5.8 &  0.3 &  116 &  02 &  14.99 &  0.02 &  3.11 & 0.07 \\ 
&5363.9282 &   3.4 &  0.3 &  178 &  03 &  15.00 &  0.02 &  3.07 & 0.06 \\ 
&5364.9272 &   8.9 &  0.4 &  151 &  02 &  15.04 &  0.02 &  2.97 & 0.07 \\ 
&5387.8569 &   6.1 &  0.3 &  186 &  02 &  14.86 &  0.02 &  3.49 & 0.08 \\ 
&5390.8325 &   3.8 &  0.3 &  154 &  02 &  14.74 &  0.02 &  3.91 & 0.09 \\ 
&5391.8589 &   3.5 &  0.4 &  162 &  03 &  14.88 &  0.02 &  3.45 & 0.08 \\ 
&5392.8271 &   3.2 &  0.4 &  150 &  04 &  14.84 &  0.02 &  3.57 & 0.08 \\ 
&5393.8359 &   4.3 &  0.4 &  156 &  03 &  14.84 &  0.01 &  3.57 & 0.05 \\ 
&5413.7930 &   5.2 &  0.4 &  164 &  02 &  14.75 &  0.01 &  3.86 & 0.05 \\ 
&5414.7808 &   2.9 &  0.4 &  155 &  04 &  14.75 &  0.01 &  3.86 & 0.05 \\ 
&5415.7773 &   3.7 &  0.4 &  156 &  03 &  14.80 &  0.01 &  3.69 & 0.05 \\ 
&5416.7695 &   4.7 &  0.4 &  161 &  02 &  14.78 &  0.01 &  3.78 & 0.05 \\ 
&5417.7744 &   5.2 &  0.4 &  166 &  02 &  14.77 &  0.01 &  3.82 & 0.05 \\ 
&5418.8096 &   4.6 &  0.4 &  171 &  02 &  14.80 &  0.02 &  3.69 & 0.08 \\ 
&5419.8135 &   8.4 &  0.5 &  191 &  02 &  14.79 &  0.02 &  3.74 & 0.09 \\ 
&5446.7095 &   5.4 &  0.5 &  151 &  02 &  14.72 &  0.01 &  4.00 & 0.05 \\ 
&5447.6626 &   2.2 &  0.5 &  137 &  04 &  14.74 &  0.02 &  3.91 & 0.09 \\ 
&5448.6694 &   1.5 &  0.8 &  155 &  10 &  14.69 &  0.01 &  4.09 & 0.06 \\ 
&5449.6602 &   1.7 &  0.8 &  162 &  08 &  14.69 &  0.01 &  4.09 & 0.06 \\ 
&5450.6621 &   2.2 &  0.7 &  153 &  07 &  14.72 &  0.02 &  4.00 & 0.09 \\ 
&5452.6562 &   7.1 &  0.9 &  196 &  05 &  14.72 &  0.01 &  4.00 & 0.06 \\ 
&5476.6138 &   5.2 &  0.6 &  182 &  04 &  14.64 &  0.02 &  4.27 & 0.09 \\ 
&5477.6196 &   4.8 &  0.4 &  180 &  02 &  14.64 &  0.01 &  4.27 & 0.06 \\ 
&5478.6240 &   6.6 &  0.6 &  162 &  03 &  14.63 &  0.01 &  4.32 & 0.06 \\ 
&5479.6304 &   8.0 &  0.5 &  154 &  02 &  14.62 &  0.01 &  4.37 & 0.06 \\ 
&5480.6123 &   2.2 &  0.4 &  203 &  04 &  14.61 &  0.01 &  4.41 & 0.06 \\ 
&5504.5923 &   6.0 &  0.5 &  143 &  02 &  14.32 &  0.01 &  5.78 & 0.08 \\ 
&5505.6133 &   3.2 &  0.5 &  160 &  03 &  14.28 &  0.01 &  5.96 & 0.08 \\ 
&5506.6655 &   3.8 &  0.6 &  158 &  03 &  14.31 &  0.01 &  5.84 & 0.09 \\ 
&5507.6108 &   4.4 &  0.5 &  162 &  02 &  14.32 &  0.01 &  5.78 & 0.08 \\ 
&5508.6191 &   3.6 &  0.5 &  170 &  03 &  14.29 &  0.01 &  5.90 & 0.08 \\ 
&5656.0156 &   2.0 &  0.7 &  158 &  08 &  14.91 &  0.02 &  3.33 & 0.08 \\ 
&5683.9736 &   2.8 &  0.6 &  124 &  04 &  14.74 &  0.02 &  3.91 & 0.09 \\ 
&5684.9956 &   2.9 &  0.7 &  142 &  05 &  14.68 &  0.01 &  4.13 & 0.06 \\ 
&5685.9941 &   1.5 &  0.8 &  101 &  12 &  14.69 &  0.02 &  4.09 & 0.09 \\ 
&5687.9917 &   3.3 &  0.7 &  134 &  05 &  14.62 &  0.01 &  4.37 & 0.06 \\ 
 \enddata
\end{deluxetable}

\begin{deluxetable}{rrcrrrrrrrrrrrr}
\tablewidth{0pt} 
\tablecaption{VARIABILITY PARAMETERS FOR 1ES~1959+650
\label{tbl-2}}
\tablehead{
\colhead{Cycle}& \colhead{Parameter}& \colhead{Average}& \colhead{Max} &
\colhead{Min}& \colhead{$\Delta_{max}$}& \colhead{$Y(\%)$}& \colhead{$\mu(\%)$}& \colhead{$\cal F$}&
\colhead{$\chi^2$} \\
\colhead{(1)}& \colhead{(2)}& \colhead{(3)}& \colhead{(4)} &
\colhead{(5)}& \colhead{(6)}& \colhead{(7)}& \colhead{(8)}&
\colhead{(9)}& \colhead{(10)} 
}
\startdata
All  & R(mag) &   14.64 $\pm$  0.24 &  15.20 &  14.08 &   1.12 &   -    &   -    &   -    &    -    \\ 
     & F(mJy) &    4.38 $\pm$  1.02 &   7.19 &   2.57 &   4.62 & 105.25 &  23.28 &   0.47 &13652.44 \\ 
     &  P(\%) &    5.7 $\pm$  2.3 &  12.2 &   1.5 &  10.7 & 187.57 &  39.87 &   0.79 & 1611.09 \\ 
     &$\theta(\degr)$&153 $\pm$ 16 & 203 & 101 & 102 &  65.60 &  10.26 &   0.34 & 3219.04 \\ 
\hline
I    & R(mag) &   14.72 $\pm$  0.07 &  14.91 &  14.60 &   0.31 &   -    &   -    &   -    &    -    \\ 
     & F(mJy) &    3.98 $\pm$  0.27 &   4.46 &   3.33 &   1.13 &  28.14 &   6.72 &   0.14 &  464.34 \\ 
     &  P(\%) &    6.9 $\pm$  2.2 &  12.2 &   3.3 &   8.9 & 127.58 &  31.87 &   0.58 &  163.13 \\ 
     &$\theta(\degr)$  &  150 $\pm$ 14 & 186 & 126 &  60 &  39.83 &   9.42 &   0.19 &  526.78 \\ 
\hline
IIa  & R(mag) &   14.23 $\pm$  0.10 &  14.40 &  14.08 &   0.32 &   -    &   -    &   -    &    -    \\ 
     & F(mJy) &    6.29 $\pm$  0.60 &   7.19 &   5.36 &   1.82 &  28.64 &   9.46 &   0.15 &  263.40 \\ 
     &  P(\%) &    5.2 $\pm$  1.0 &   7.3 &   3.8 &   3.5 &  61.49 &  19.22 &   0.32 &   45.28 \\ 
     &$\theta(\degr)$  &  149 $\pm$  6 & 159 & 138 &  21 &  13.47 &   4.17 &   0.07 &   73.07 \\ 
\hline
IIb  & R(mag) &   14.55 $\pm$  0.10 &  14.74 &  14.42 &   0.32 &   -    &   -    &   -    &    -    \\ 
     & F(mJy) &    4.67 $\pm$  0.43 &   5.25 &   3.91 &   1.34 &  28.35 &   9.24 &   0.15 &  598.68 \\ 
     &  P(\%) &    7.6 $\pm$  1.4 &  10.3 &   5.3 &   5.0 &  63.52 &  18.78 &   0.32 &  138.64 \\ 
     &$\theta(\degr)$  &  151 $\pm$ 11 & 171 & 136 &  35 &  23.10 &   7.29 &   0.11 &  563.81 \\ 
\hline
III  & R(mag) &   14.77 $\pm$  0.23 &  15.20 &  14.28 &   0.92 &   -    &   -    &   -    &    -    \\ 
     & F(mJy) &    3.90 $\pm$  0.88 &   5.96 &   2.57 &   3.39 &  86.83 &  22.52 &   0.40 & 6005.49 \\ 
     &  P(\%) &    4.1 $\pm$  1.8 &   8.9 &   1.5 &   7.4 & 176.22 &  43.51 &   0.72 &  553.05 \\ 
     &$\theta(\degr)$  &  157 $\pm$ 19 & 203 & 101 & 102 &  64.01 &  12.39 &   0.34 & 1657.01 \\
\enddata
\tablecomments{There are no statistics $ Y, \mu, \cal F$ and $ \chi^2$ for the
  magnitude due to its logarithmic character.}

\end{deluxetable}

\begin{deluxetable}{lrrrrcrrrrrr}
\tablewidth{0pt} \tablecaption{POLARIZATION FOR THE VARIABLE COMPONENT OF 1ES~1959+650 IN CYCLES IIa and IIb.
\label{tbl-3}}
\tablehead{
\colhead{Cycle}&\colhead{$q_{var}$}&\colhead{$r_q$} &
\colhead{$u_{var}$}&\colhead{$r_u$}&\colhead{$p_{var}(\%)$}&\colhead{$\theta_{var}(
  ^\circ)$}\\
\colhead{(1)}&\colhead{(2)}&\colhead{(3)} &
\colhead{(4)}&\colhead{(5)}&\colhead{(6)}&\colhead{(7)}
}
\startdata
IIa     &  0.228 $\pm$  0.028 & 0.913 & -0.408 $\pm$ 0.054 & -0.902 & 46.8 $\pm$ 4.9 & 150 $\pm$ 4\\
IIb     &  0.304 $\pm$  0.044 & 0.872 &  0.275 $\pm$ 0.015 &  0.980 & 41.0 $\pm$ 3.4 &  21 $\pm$ 4\\
\enddata

\end{deluxetable}

\begin{deluxetable}{lrrrrcrrrrr}
\tablewidth{0pt} \tablecaption{THE VARIABLE COMPONENT FOR TWO-COMPONENT MODEL\label{tbl-4}}
\tablehead{ 
\colhead{JD}& \colhead{$p_{var}$}&
\colhead{$\theta_{var}$}&  \colhead{$I_{var}$} \\
\colhead{2450000.00+}& \colhead{($\%$)}&
\colhead{(\,$^{\circ}\,$)}& \colhead{(mJy)} \\
\colhead{(1)}& \colhead{(2)}& \colhead{(3)}& \colhead{(4)}
}
\startdata  
4591.9907 &   6.8 $\pm$  0.7 &  145 $\pm$ 05 &  3.00 $\pm$  0.26 \\ 
4593.9912 &   7.6 $\pm$  1.1 &  134 $\pm$ 06 &  2.35 $\pm$  0.29 \\ 
4594.9883 &   8.4 $\pm$  1.1 &  131 $\pm$ 05 &  2.57 $\pm$  0.29 \\ 
4620.9575 &   3.8 $\pm$  0.5 &  140 $\pm$ 10 &  2.62 $\pm$  0.29 \\ 
4621.9854 &   4.9 $\pm$  0.7 &  139 $\pm$ 11 &  2.27 $\pm$  0.29 \\ 
4622.9287 &   3.0 $\pm$  0.3 &  167 $\pm$ 14 &  2.57 $\pm$  0.25 \\ 
4623.9331 &   4.7 $\pm$  0.5 &  163 $\pm$ 10 &  2.62 $\pm$  0.25 \\ 
4624.9839 &   9.8 $\pm$  1.3 &  167 $\pm$ 05 &  2.27 $\pm$  0.25 \\ 
4655.9331 &  13.4 $\pm$  1.6 &  164 $\pm$ 03 &  2.44 $\pm$  0.25 \\ 
4656.9160 &  17.5 $\pm$  2.2 &  141 $\pm$ 03 &  2.31 $\pm$  0.25 \\ 
4660.9390 &   8.6 $\pm$  1.0 &  151 $\pm$ 04 &  2.48 $\pm$  0.25 \\ 
4661.8857 &   8.0 $\pm$  1.0 &  154 $\pm$ 06 &  2.35 $\pm$  0.25 \\ 
4681.9053 &  12.2 $\pm$  1.4 &  186 $\pm$ 04 &  2.53 $\pm$  0.25 \\ 
4683.8711 &   4.6 $\pm$  0.5 &  151 $\pm$ 09 &  2.48 $\pm$  0.25 \\ 
4707.7686 &  12.5 $\pm$  1.7 &  129 $\pm$ 05 &  2.57 $\pm$  0.29 \\ 
4712.7568 &   7.5 $\pm$  0.8 &  133 $\pm$ 07 &  3.00 $\pm$  0.26 \\ 
4737.6260 &   5.4 $\pm$  0.6 &  155 $\pm$ 07 &  2.66 $\pm$  0.25 \\ 
4738.6494 &   7.4 $\pm$  0.8 &  173 $\pm$ 05 &  2.76 $\pm$  0.26 \\ 
4764.6245 &  11.5 $\pm$  1.4 &  140 $\pm$ 04 &  2.80 $\pm$  0.29 \\ 
4766.6401 &   7.9 $\pm$  0.8 &  147 $\pm$ 06 &  2.85 $\pm$  0.26 \\ 
4767.6763 &  10.2 $\pm$  1.1 &  155 $\pm$ 05 &  2.85 $\pm$  0.26 \\ 
4768.6421 &   7.7 $\pm$  0.9 &  140 $\pm$ 06 &  2.66 $\pm$  0.25 \\ 
4771.6611 &  15.0 $\pm$  1.9 &  122 $\pm$ 04 &  2.35 $\pm$  0.25 \\ 
4773.6025 &  15.9 $\pm$  2.0 &  190 $\pm$ 03 &  2.40 $\pm$  0.25 \\ 
4774.6011 &  10.5 $\pm$  1.4 &  136 $\pm$ 06 &  2.14 $\pm$  0.24 \\ 
4803.6006 &   5.8 $\pm$  0.8 &  154 $\pm$ 09 &  2.14 $\pm$  0.25 \\ 
4804.5962 &   7.8 $\pm$  1.3 &  132 $\pm$ 09 &  1.87 $\pm$  0.26 \\ 
4944.9717 &   8.5 $\pm$  0.6 &  143 $\pm$ 04 &  4.19 $\pm$  0.27 \\ 
4973.9253 &   6.4 $\pm$  0.5 &  141 $\pm$ 05 &  4.37 $\pm$  0.28 \\ 
4974.9087 &   3.7 $\pm$  0.3 &  143 $\pm$ 08 &  4.62 $\pm$  0.28 \\ 
4975.9033 &   4.2 $\pm$  0.3 &  134 $\pm$ 06 &  4.49 $\pm$  0.28 \\ 
4976.8965 &   4.2 $\pm$  0.3 &  139 $\pm$ 06 &  4.37 $\pm$  0.27 \\ 
4977.9121 &   5.2 $\pm$  0.4 &  143 $\pm$ 04 &  4.25 $\pm$  0.27 \\ 
4979.8735 &   5.2 $\pm$  0.5 &  145 $\pm$ 04 &  3.89 $\pm$  0.31 \\ 
5002.9106 &   7.2 $\pm$  0.5 &  158 $\pm$ 03 &  5.72 $\pm$  0.36 \\ 
5003.8867 &   6.1 $\pm$  0.4 &  153 $\pm$ 03 &  5.50 $\pm$  0.36 \\ 
5004.9136 &   6.2 $\pm$  0.5 &  153 $\pm$ 03 &  5.50 $\pm$  0.36 \\ 
5005.8696 &   6.0 $\pm$  0.4 &  151 $\pm$ 03 &  5.22 $\pm$  0.35 \\ 
5006.8984 &   6.3 $\pm$  0.5 &  161 $\pm$ 03 &  4.75 $\pm$  0.36 \\ 
5007.8599 &   5.1 $\pm$  0.4 &  151 $\pm$ 04 &  5.15 $\pm$  0.36 \\ 
5008.8726 &   4.3 $\pm$  0.3 &  155 $\pm$ 04 &  5.50 $\pm$  0.36 \\ 
5059.7646 &  12.5 $\pm$  1.2 &  145 $\pm$ 03 &  3.67 $\pm$  0.31 \\ 
5060.7266 &  11.5 $\pm$  1.1 &  146 $\pm$ 02 &  3.78 $\pm$  0.31 \\ 
5062.7378 &  12.7 $\pm$  1.2 &  154 $\pm$ 02 &  3.67 $\pm$  0.31 \\ 
5063.7241 &  11.4 $\pm$  1.1 &  146 $\pm$ 03 &  3.73 $\pm$  0.31 \\ 
5064.7183 &  11.4 $\pm$  1.1 &  148 $\pm$ 02 &  3.62 $\pm$  0.31 \\ 
5092.6494 &   9.3 $\pm$  0.8 &  137 $\pm$ 03 &  3.45 $\pm$  0.26 \\ 
5093.6284 &   9.3 $\pm$  0.8 &  137 $\pm$ 03 &  3.51 $\pm$  0.26 \\ 
5094.6235 &   9.2 $\pm$  0.9 &  135 $\pm$ 03 &  3.14 $\pm$  0.26 \\ 
5095.6250 &   8.2 $\pm$  0.8 &  133 $\pm$ 04 &  3.19 $\pm$  0.26 \\ 
5096.6221 &   8.3 $\pm$  0.8 &  136 $\pm$ 04 &  3.14 $\pm$  0.26 \\ 
5097.6221 &   8.6 $\pm$  0.8 &  144 $\pm$ 03 &  3.25 $\pm$  0.26 \\ 
5098.6680 &   9.5 $\pm$  0.9 &  143 $\pm$ 03 &  3.14 $\pm$  0.26 \\ 
5122.6343 &   7.6 $\pm$  0.8 &  156 $\pm$ 04 &  3.00 $\pm$  0.29 \\ 
5123.6606 &   8.3 $\pm$  0.8 &  154 $\pm$ 04 &  3.14 $\pm$  0.26 \\ 
5124.6235 &   9.0 $\pm$  0.8 &  166 $\pm$ 03 &  3.40 $\pm$  0.26 \\ 
5150.6030 &   9.6 $\pm$  1.3 &  176 $\pm$ 03 &  2.44 $\pm$  0.28 \\ 
5152.5942 &   6.5 $\pm$  0.7 &  173 $\pm$ 04 &  2.53 $\pm$  0.25 \\ 
5153.5879 &   7.5 $\pm$  0.8 &  168 $\pm$ 04 &  2.57 $\pm$  0.25 \\ 
5154.5879 &   7.2 $\pm$  0.8 &  168 $\pm$ 04 &  2.53 $\pm$  0.25 \\ 
 \enddata
\end{deluxetable}

\begin{deluxetable}{lcrrrrl}
\tablewidth{0pt} \tablecaption{PHYSICAL PARAMETERS FOR 1ES~1959+650
 \label{tbl-5}}
\tablehead{\colhead{Parameter}&\colhead{Value}&\colhead{Units}}
\startdata
 $B  $        & ~0.06  $\pm$ 0.01 & Gauss      &\\ 
 $\delta_0$   & 23.3   $\pm$ 0.4  &            &\\
 $\Psi_0$     & 72.9   $\pm$ 6.1  & degree     &\\
 $\Phi_0$     &  2.35  $\pm$ 0.26 & degree     &\\
 $r_b$        &  5.61  $\pm$ 0.68 &$10^{17}$ cm &\\
 $r_s$        &  1.29  $\pm$ 0.26 &$10^{17}$ cm &\\
 $l_B$        &  1.31  $\pm$ 0.16 &$10^{17}$ cm &\\
 $t_{min}$    &  9.74  $\pm$ 1.17 & days &\\
  \enddata
\tablecomments{The parameters were calculated using values derived by
 \citet{2004ApJ...600..115P} and \citet{2008ApJ...678...64P,2010ApJ...723.1150P} 
 for the shock speed $\beta_s=0.1$, and the Lorentz factor of the shock $\Gamma_s\,\sim$3. In addition, we use the value
  for the optical spectral index $\alpha_{ox}= 1.64$, reported by \citet{2010ApJ...716...30A}; and the value of the bulk Lorentz factor of the jet $\Gamma_j$=18 from \citet{2008ApJ...679.1029T}. 
}
\end{deluxetable}

\begin{deluxetable}{lrrrrcrrrrr}
\tablewidth{0pt} \tablecaption{VARIABLE PHYSICAL PARAMETERS FROM POLARIMETRIC DATA OBTAINED FOR 1ES~1959+650
\label{tbl-6}}
\tablehead{ 
\colhead{JD}& \colhead{$\delta(t)$}&
\colhead{$\Phi(t)$}&  \colhead{$\Psi(t)$}& \colhead{$\eta(t)$}\\
\colhead{2450000+}& &
\colhead{[\,$^{\circ}\,$]}& \colhead{[\,$^{\circ}\,$]}&  & \\
\colhead{(1)}& \colhead{(2)}& \colhead{(3)}& \colhead{(4)}& \colhead{(5)}
}
\startdata  
4944.9717 &  22.1 $\pm$  0.4 & 2.52 $\pm$ 0.06 &  76.8 $\pm$  1.3 & 1.102 $\pm$ 0.008 \\ 
4973.9253 &  22.3 $\pm$  0.4 & 2.50 $\pm$ 0.06 &  76.3 $\pm$  1.3 & 1.080 $\pm$ 0.011 \\ 
4974.9087 &  22.5 $\pm$  0.4 & 2.47 $\pm$ 0.06 &  75.6 $\pm$  1.3 & 1.052 $\pm$ 0.012 \\ 
4975.9033 &  22.4 $\pm$  0.4 & 2.49 $\pm$ 0.06 &  76.0 $\pm$  1.3 & 1.056 $\pm$ 0.011 \\ 
4976.8965 &  22.3 $\pm$  0.4 & 2.50 $\pm$ 0.06 &  76.3 $\pm$  1.3 & 1.056 $\pm$ 0.010 \\ 
4977.9121 &  22.2 $\pm$  0.4 & 2.52 $\pm$ 0.06 &  76.6 $\pm$  1.3 & 1.067 $\pm$ 0.005 \\ 
4979.8735 &  21.9 $\pm$  0.4 & 2.56 $\pm$ 0.06 &  77.6 $\pm$  1.3 & 1.067 $\pm$ 0.006 \\ 
5002.9106 &  23.3 $\pm$  0.4 & 2.35 $\pm$ 0.06 &  72.9 $\pm$  1.4 & 1.095 $\pm$ 0.007 \\ 
5003.8867 &  23.1 $\pm$  0.4 & 2.38 $\pm$ 0.06 &  73.4 $\pm$  1.4 & 1.081 $\pm$ 0.006 \\ 
5004.9136 &  23.1 $\pm$  0.4 & 2.38 $\pm$ 0.06 &  73.4 $\pm$  1.4 & 1.083 $\pm$ 0.007 \\ 
5005.8696 &  22.9 $\pm$  0.4 & 2.41 $\pm$ 0.06 &  74.1 $\pm$  1.4 & 1.079 $\pm$ 0.006 \\ 
5006.8984 &  22.6 $\pm$  0.4 & 2.46 $\pm$ 0.06 &  75.3 $\pm$  1.3 & 1.080 $\pm$ 0.007 \\ 
5007.8599 &  22.9 $\pm$  0.4 & 2.41 $\pm$ 0.06 &  74.3 $\pm$  1.4 & 1.068 $\pm$ 0.006 \\ 
5008.8726 &  23.1 $\pm$  0.4 & 2.38 $\pm$ 0.06 &  73.4 $\pm$  1.4 & 1.060 $\pm$ 0.006 \\ 
5059.7646 &  21.7 $\pm$  0.4 & 2.59 $\pm$ 0.06 &  78.3 $\pm$  1.3 & 1.142 $\pm$ 0.008 \\ 
5060.7266 &  21.8 $\pm$  0.4 & 2.58 $\pm$ 0.06 &  78.0 $\pm$  1.3 & 1.133 $\pm$ 0.007 \\ 
5062.7378 &  21.7 $\pm$  0.4 & 2.59 $\pm$ 0.06 &  78.3 $\pm$  1.3 & 1.145 $\pm$ 0.007 \\ 
5063.7241 &  21.7 $\pm$  0.4 & 2.58 $\pm$ 0.06 &  78.1 $\pm$  1.3 & 1.132 $\pm$ 0.007 \\ 
5064.7183 &  21.6 $\pm$  0.4 & 2.60 $\pm$ 0.06 &  78.4 $\pm$  1.2 & 1.130 $\pm$ 0.007 \\ 
5092.6494 &  21.5 $\pm$  0.4 & 2.62 $\pm$ 0.06 &  78.9 $\pm$  1.2 & 1.105 $\pm$ 0.008 \\ 
5093.6284 &  21.5 $\pm$  0.4 & 2.61 $\pm$ 0.06 &  78.8 $\pm$  1.2 & 1.106 $\pm$ 0.008 \\ 
5094.6235 &  21.2 $\pm$  0.4 & 2.67 $\pm$ 0.06 &  79.9 $\pm$  1.2 & 1.101 $\pm$ 0.008 \\ 
5095.6250 &  21.2 $\pm$  0.4 & 2.66 $\pm$ 0.06 &  79.7 $\pm$  1.2 & 1.091 $\pm$ 0.008 \\ 
5096.6221 &  21.2 $\pm$  0.4 & 2.67 $\pm$ 0.06 &  79.9 $\pm$  1.2 & 1.093 $\pm$ 0.008 \\ 
5097.6221 &  21.3 $\pm$  0.4 & 2.65 $\pm$ 0.06 &  79.6 $\pm$  1.2 & 1.098 $\pm$ 0.008 \\ 
5098.6680 &  21.2 $\pm$  0.4 & 2.67 $\pm$ 0.06 &  79.9 $\pm$  1.2 & 1.105 $\pm$ 0.008 \\ 
5122.6343 &  21.0 $\pm$  0.4 & 2.69 $\pm$ 0.06 &  80.4 $\pm$  1.2 & 1.087 $\pm$ 0.008 \\ 
5123.6606 &  21.2 $\pm$  0.4 & 2.67 $\pm$ 0.06 &  79.9 $\pm$  1.2 & 1.094 $\pm$ 0.007 \\ 
5124.6235 &  21.4 $\pm$  0.4 & 2.63 $\pm$ 0.06 &  79.1 $\pm$  1.2 & 1.101 $\pm$ 0.008 \\ 
5150.6030 &  20.4 $\pm$  0.4 & 2.78 $\pm$ 0.06 &  82.3 $\pm$  1.2 & 1.095 $\pm$ 0.009 \\ 
5152.5942 &  20.5 $\pm$  0.4 & 2.77 $\pm$ 0.06 &  82.0 $\pm$  1.1 & 1.070 $\pm$ 0.008 \\ 
5153.5879 &  20.6 $\pm$  0.4 & 2.76 $\pm$ 0.06 &  81.8 $\pm$  1.1 & 1.081 $\pm$ 0.008 \\ 
5154.5879 &  20.5 $\pm$  0.4 & 2.77 $\pm$ 0.06 &  82.0 $\pm$  1.1 & 1.078 $\pm$ 0.008 \\ 
\enddata
\tablecomments{ Active cycle 2009 for 1ES 1959+650. Col.~(1): Julian
  Day; Col.~(2): Doppler factor; Col.~(3): Angle between the line of
  sight and jet axis; Col.~(4): Viewing angle of the shock; Col.~(5):
  Ratio of the density in the shocked region with respect to unshocked
  region.}
\end{deluxetable}

\end{document}